\documentclass[9pt,twocolumn,twoside]{optica}
\usepackage{xspace} 
\usepackage{bm}

\setboolean{shortarticle}{false}
\setboolean{minireview}{true}

\newcommand{\vp}{v_{\text{p}}}
\newcommand{\vg}{v_{\text{g}}}
\newcommand{\vm}{v_{\text{m}}}
\newcommand{\vs}{v_{\text{s}}}
\newcommand{\Vb}{V_{\text{b}}}

\newcommand{\kappam}{\gamma}
\newcommand{\betap}{\beta_{\text{p}}}
\newcommand{\omegam}{\omega_{\text{m}}}
\newcommand{\omegao}{\omega_{\text{o}}}
\newcommand{\omegain}{\omega_{\text{in}}}

\newcommand{\neff}{n_{\text{eff}}}

\newcommand{\Qm}{Q_{\text{m}}}
\newcommand{\Cm}{C_{\text{m}}}
\newcommand{\Cp}{C_{\text{p}}}
\newcommand{\omegamu}{\omega_{\mu}}

\newcommand{\keff}{k_{\text{eff}}}

\newcommand{\partialx}{\partial_{x}}
\newcommand{\partialV}{\partial_{V}}

\newcommand{\kappaex}{\kappa_\text{ex}}
\newcommand{\kappain}{\kappa_\text{in}}
\newcommand{\gammain}{\gamma_\text{in}}
\newcommand{\alpham}{\alpha_{\text{m}}}
\newcommand{\etap}{\eta_{\text{p}}}
\newcommand{\xzp}{x_\text{zp}}
\newcommand{\xpi}{x_{\pi}}
\newcommand{\Vpi}{V_{\pi}}
\newcommand{\Vzp}{V_{\text{zp}}}
\newcommand{\meff}{m_\text{eff}}
\newcommand{\hairsp}{\hspace{1pt}} 
\newcommand{\eg}{\textit{e.\hairsp{}g.}\xspace} 
\newcommand*\mean[1]{\overline{#1}}
\newcommand{\dyad}[1]{\mbox{$\overline{\overline{#1}}$}}
\newcommand{\curl}{\mbox{$\text{curl}~$}}

\newcommand{\fr}{\mbox{$(\m r)$}}

\newcommand{\ket}[1]{\mbox{$|#1\rangle$}}
\newcommand{\bra}[1]{\mbox{$\langle#1|$}}

\newcommand{\op}[2]{\mbox{$\hat{#1}_{#2}$}}

\newcommand{\m}[1]{\mbox{$\mathbf{#1}$}}
\def\be{\begin{equation}}
\def\ee{\end{equation}}

\title{Controlling phonons and photons at the wavelength-scale: \\ silicon photonics meets silicon phononics}

\author[1]{Amir H. Safavi-Naeini}
\author[2]{Dries Van Thourhout}
\author[2]{Roel Baets}
\author[1,2]{Rapha\"{e}l Van Laer}
\affil[1]{Department of Applied Physics and Ginzton Laboratory, Stanford University, USA}
\affil[2]{Photonics Research Group (INTEC), Department of Information Technology, Ghent University--imec, Belgium \newline
\textcolor{white}{.}Center for Nano- and Biophotonics, Ghent University, Belgium}

\dates{Compiled \today}

\ociscodes{}

\begin{abstract}
Radio-frequency communication systems have long used bulk- and surface-acoustic-wave devices supporting ultrasonic mechanical waves to manipulate and sense signals. These devices have greatly improved our ability to process microwaves by interfacing them to orders-of-magnitude slower and lower loss mechanical fields. In parallel, long-distance communications have been dominated by low-loss infrared optical photons. As electrical signal processing and transmission approaches physical limits imposed by energy dissipation, optical links are now being actively considered for mobile and cloud technologies. Thus there is a strong driver for wavelength-scale mechanical wave or ``phononic'' circuitry fabricated by scalable semiconductor processes. With the advent of these circuits, new micro- and nanostructures that combine electrical, optical and mechanical elements have emerged. In these devices, such as optomechanical waveguides and resonators, optical photons and gigahertz phonons are ideally matched to one another as both have wavelengths on the order of micrometers. The development of phononic circuits has thus emerged as a vibrant field of research pursued  for  optical signal processing and sensing applications as well as emerging quantum technologies. In this review, we discuss the key physics and figures of merit underpinning this field. We also summarize the state of the art in nanoscale electro- and optomechanical systems with a focus on scalable platforms such as silicon. Finally, we give perspectives on what these new systems may bring and what challenges they face in the coming years. In particular, we believe hybrid electro- and optomechanical devices incorporating highly coherent and compact mechanical elements on a chip have significant untapped potential for electro-optic modulation, quantum microwave-to-optical photon conversion, sensing and microwave signal processing.
\end{abstract}

\setboolean{displaycopyright}{false}

\begin{document}

\maketitle
\tableofcontents

\section{Introduction}

Microwave-frequency acoustic or mechanical wave devices have found numerous applications in radio-signal processing and sensing. They already form mature technologies with large markets. The vast majority of these devices are made of piezoelectric materials that are driven by electrical circuits \cite{Hashimoto2000,Hashimoto2009a,Jones2013,Ruby2001,Ruby2007,Hara2005}. A major technical challenge in such systems is obtaining the suitable matching conditions for efficient conversion between electrical and mechanical energy. Typically, this entails reducing the effective electrical impedance of the electromechanical component by increasing the capacitance of the driving element. This has generally led to devices with large capacitors that drive mechanical modes with large mode volumes. Here, we describe a recent shift in research towards structures that are only about a wavelength, i.e. roughly one micron at gigahertz frequencies, across in two or more dimensions.

\begin{figure}[ht]
\centering
\includegraphics[width=\linewidth]{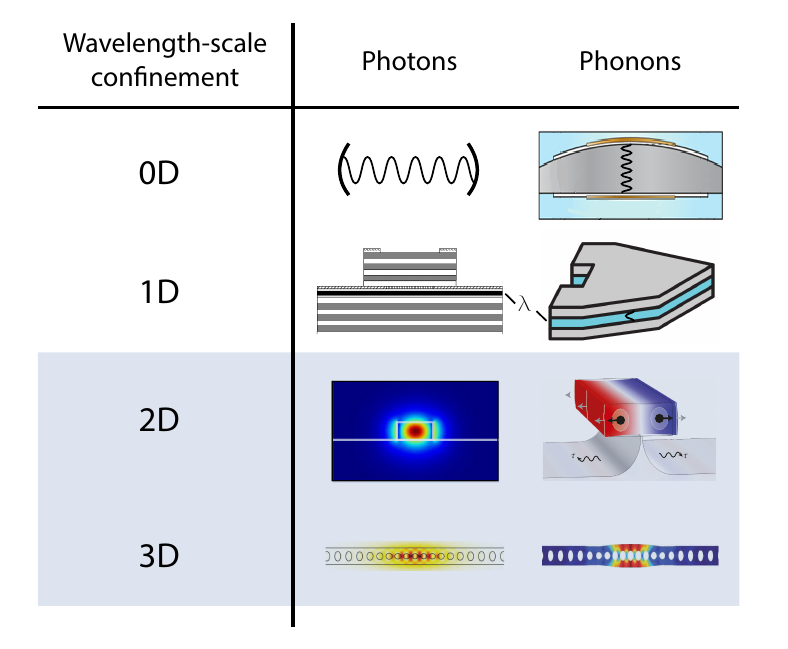}
\caption{\textbf{Confining photons and phonons to the wavelength-scale.} Photonic and phononic systems can be classified according to the number of dimensions in which they confine photons and phonons to a wavelength. New structures have emerged in which both photons and phonons are confined to the wavelength-scale in two or three spatial dimensions. Here we focus on such 2D- or 3D-confined wavelength-scale systems at gigahertz frequencies. Related reviews on integrated opto- and electromechanical systems are  \cite{VanThourhout2010a,Eggleton2013,Aspelmeyer2014,Merklein2016,Midolo2018a,Ekinci2005NanoelectromechanicalSystems}. The table gives as examples of 2D- and 3D-confined devices a sub-$\mu\textrm{m}^{2}$ silicon photonic-phononic waveguide \cite{VanLaer2015} and a sub-$\mu\textrm{m}^{3}$ silicon optomechanical crystal \cite{Safavi-Naeini2012}.  The depicted 0D- and 1D-confined structures are a long Fabry-P\'{e}rot cavity, a vertical-cavity surface-emitting laser \cite{Passaro2005} for the optical case and a thick quartz \cite{Goryachev2012a} and a thin aluminum nitride BAW resonator \cite{OConnell2010} for the mechanical case.}
\label{fig:table_of_confinements}
\end{figure}

Greater confinement of mechanical waves in a device has both advantages and drawbacks depending on the application at hand. In the case of interactions with optical fields, higher confinement increases the strength and speed of the interaction allowing faster switching and lower powers. A smaller system demands less dissipated energy to achieve the same effects, simply because it focuses all of the optical and mechanical energy into a smaller volume. High confinement also enables scalable, less costly fabrication with more functionality packed into a smaller space. Perhaps more importantly, in analogy to microwave and photonic circuits that become significantly easier to engineer in the single- and few-moded limits, obtaining control over the full mode structure of the devices vastly simplifies designing and scaling systems to higher complexity. Confining mechanical energy is not without its drawbacks; as we will see below, focusing the mechanical energy into a small volume also means that deleterious nonlinear effects manifest at lower powers, and matching directly to microwave circuits becomes significantly more difficult due to vanishing capacitances. We can classify confinement in terms of its dimensionality [Fig. \ref{fig:table_of_confinements}]. The dimensionality refers to the number of dimensions where confinement is on the scale of the wavelength of the excitation in bulk. For example, surface acoustic wave (SAW) resonators \cite{Hashimoto2000}, much like thin-film bulk acoustic wave (BAW) resonators \cite{Hashimoto2009a}, have wavelength-scale confinement in only one dimension -- perpendicular to the chip surface -- and are therefore 1D-confined. Until a few years ago, wavelength-scale phononic confinement at gigahertz frequencies beyond 1D remained out of reach.

Intriguingly, both near-infrared optical photons and gigahertz phonons have a wavelength of about one micron. This results from the five orders of magnitude difference in the speed of light relative to the speed of sound. The fortuitous matching of length scales was used to demonstrate the first 2D- and 3D-confined systems in which both photons and phonons are confined to the same area or volume [Fig. \ref{fig:table_of_confinements}]. These measurements have been enabled by advances in low-loss photonic circuits that couple light to material deformations through boundary and photoelastic perturbations. Direct capacitive or piezoelectric coupling to these types of resonances has been harder since the relatively low speed of sound in solid-state materials means that gigahertz-frequency phonons have very small volume, leading to miniscule electrically-induced forces at reasonable voltages, or in other words large motional resistances that are difficult to match to standard microwave circuits \cite{Nguyen2013MEMS-basedCommunications}.  

In this review, we primarily consider recent advances in gigahertz-frequency phononic devices. These devices have been demonstrated mainly in the context of photonic circuits, and share many commonalities with integrated photonic structures in terms of their design and physics. They also have the potential to realize important new functionalities in photonic circuits. Despite recent demonstrations of confined phonon devices operating at gigahertz frequencies and coupled to optical fields, phononic circuits are still in their infancy, and applications beyond those of interest in integrated photonics remain largely unexplored. Several attractive aspects of mechanical elements remain unrealized in chip-scale systems, especially in those based on non-piezoelectric materials. In this review, we first describe the basic physics underpinning this field with specific attention to the mechanical aspects of optomechanical devices. We discuss common approaches used to guide and confine mechanical waves in nanoscale structures in section \ref{sec:guideconfine}. Next, we describe the key mechanisms behind interactions between phonons and both optical and microwave photons in section \ref{sec:photonphononinteractions}. These interactions allow us to efficiently generate and read-out mechanical waves on a chip. Section \ref{sec:stateoftheart} briefly summarizes the state of the art in opto- and electromechanical devices. It also describes a few commonly used figures of merit in this field. Finally, we give our perspectives on the field in section \ref{sec:perspectives}. In analogy to silicon photonics \cite{Miller2017b,Wang2017b,Sun2015,Rahim2016,Atabaki2018,Soref2006b}, the field may be termed ``silicon phononics''. While not strictly limited to the material silicon, its goal is to develop a platform whose fabrication is in principle scalable to many densely integrated mechanical devices.

\section{Guiding and confining phonons}
\label{sec:guideconfine}
Phonons obey broadly similar physics as photons so they can be guided and confined by comparable mechanisms, as detailed in the following subsections.

\subsection{Total internal reflection}
\label{subsec:TIR}
\begin{figure}[ht]
\centering
\includegraphics[width = \linewidth]{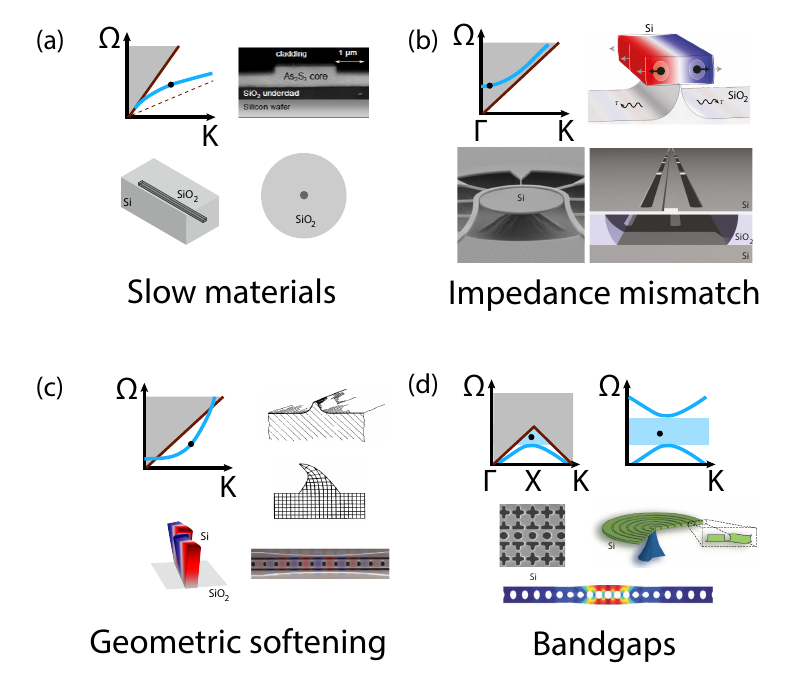}
\caption{\textbf{Mechanisms for phononic confinement in micro- and nanostructures.} We illustrate the main approaches with phononic dispersion diagrams $\Omega(K)$ and mark the operating point in black. (a) A waveguide core whose mechanical excitations propagate slower than the slowest waves in the surrounding materials supports acoustic total internal reflection. Examples include chalcogenide rib waveguides on silica \cite{Eggleton2013}, silica waveguides cladded by silicon \cite{Liu2017} and Ge-doped fibers \cite{Kobyakov2009}. (b) Even when phonons are phase-matched to surface- or bulk-excitations, their leakage can be limited by impedance mismatch such as in suspended silicon waveguides and disks \cite{VanLaer2015,VanLaer2015c,Shin2013b,Kittlaus2016,Sun2012} and silica microtoroids \cite{Verhagen2012}. (c) In contrast to optical fields, mechanical waves can be trapped by surface perturbations that soften the elastic response such as in case of Rayleigh surface waves \cite{Hashimoto2000}, silicon fins on silica \cite{Sarabalis2016,Sarabalis2017a} and all-silicon surface perturbations \cite{Oliner1973,Ash1969,Lagasse,Lagasse1973}. (d) Finally, phonons can be trapped to line- or point-defects in periodic structures with a phononic bandgap such as in silicon optomechanical crystals \cite{Safavi-Naeini2012,Safavi-naeini2014}, line defects \cite{Fang2016} and bullseye disks \cite{Santos2017}. Many structures harness a combination of these mechanisms.}
\label{fig:guidingphonons}
\end{figure}

In a system with continuous translational symmetry, waves incident on a medium totally reflect when they are not phase-matched to any excitations in that medium. This is called total internal reflection. The waves can be confined inside a slow medium sandwiched between two faster media by this mechanism [Fig. \ref{fig:guidingphonons}a]. This ensures that at fixed frequency $\Omega$ the guided wave is not phase-matched to any leaky waves since its wavevector $K(\Omega) = \Omega/v_{\phi}$ -- with $v_{\phi} $ its phase velocity -- exceeds the largest wavevector among waves in the surrounding media at that frequency. In other words, the confined waves must have maximal slowness $1/v_{\phi}$. This principle applies to both optical and mechanical fields \cite{Auld1973,Saleh1991}.

Still, there are important differences between the optical and mechanical cases. For instance, a bulk material has only two transverse optical polarizations while it sustains two transverse mechanical polarizations with speed $v_{\text{t}}$ and a longitudinally polarized mechanical wave with speed $v_{\text{l}}$. Unlike in the optical case, these polarizations generally mix in a complex way at interfaces \cite{Auld1973}. In addition, a boundary between a material and air leads to geometric softening (see next section), a situation in which interfaces reduce the speed of certain mechanical polarizations. This generates slow SAW modes that are absent in the optical case. So achieving mechanical confinement requires care in looking for the slowest waves in the surrounding structures. These are often surface instead of bulk excitations. Among the bulk excitations, transversely polarized are slower than longitudinally polarized phonons ($v_{\text{t}} < v_{\text{l}}$).

Conflicting demands often arise when designing waveguides or cavities to confine photons and phonons in the same region: photons can be confined easily in dense media with a high refractive index and thus small speed of light but phonons are naturally trapped in soft and light materials with a small speed of sound. In particular, the mechanical phase velocities scale as $v_{\phi} = \sqrt{E/\rho}$ with $E$ the stiffness or Young's modulus and $\rho$ the mass density. For instance, a waveguide core made of silicon (refractive index $n_{\text{Si}} = 3.5$) and embedded in silica ($n_{\text{SiO}_{2}} = 1.45$) strongly confines photons by total internal reflection but cannot easily trap phonons (for exceptions see next sections). On the other hand, a waveguide core made of silica ($v_{\text{t}} = 5500 \, \text{m/s}$) embedded in silicon ($v_{\text{t}} = 5843 \, \text{m/s}$) can certainly trap mechanical \cite{Liu2017} but not optical fields. Still, some structures find a sweet spot in this trade-off: the principle of total internal reflection is currently exploited to guide phonons in Ge-doped optical fibers \cite{Kobyakov2009} and chalcogenide waveguides \cite{Eggleton2013}.

Since silicon is ``slower'' than silicon dioxide optically, but ``faster'' acoustically, simple index guiding for co-confined optical and mechanical fields is not an option in the canonical platform of silicon photonics, silicon-on-insulator. Below we consider techniques that circumvent this limitation and enable strongly co-localized optomechanical waves and interactions.

\subsection{Impedance mismatch}
\label{subsec:impedancemismatch}
The generally conflicting demands between photonic and phononic confinement (see above) can be reconciled through impedance mismatch [Fig. \ref{fig:guidingphonons}c]. The characteristic acoustic impedance of a medium is $Z_{\text{m}} = \rho v_{\phi}$ with $\rho$ the mass density \cite{Auld1973}. Interfaces between media with widely different impedances $Z_{\text{m}}$, such as between solids and gases, strongly reflect phonons. In addition, gases have an acoustic cut-off frequency -- set by the molecular mean-free path -- above which they do not support acoustic excitations \cite{Royer2000}. At atmospheric pressure this frequency is roughly $\Omega_\text{c}/(2\pi) \approx 0.5 \, \text{GHz}$. Above this frequency $\Omega_\text{c}$ acoustic leakage and damping because of air are typically negligible. The cut-off frequency $\Omega_\text{c}$ can be drastically reduced with vacuum chambers, an approach that has been pursued widely to confine low-frequency phonons \cite{Aspelmeyer2014}. These ideas were harnessed in silicon-on-insulator waveguides to confine both photons and phonons to silicon waveguide cores \cite{VanLaer2015,VanLaer2015c,Kittlaus2016} over milli- to centimeter propagation lengths. The acoustic impedances of silicon and silica are quite similar, so in these systems the silica needs to be removed to realize low phonon leakage from the silicon core. In one approach \cite{VanLaer2015}, the silicon waveguide was partially under-etched to leave a small silica pillar that supports the waveguide [Fig. \ref{fig:guidingphonons}c]. In another, the silicon waveguide was fully suspended while leaving periodic silica or silicon anchors \cite{VanLaer2015c,Kittlaus2016}.

\subsection{Geometric softening}
\label{subsec:geomsoftening}
The guided wave structures considered above utilize full or partial under-etching of the oxide layer to prevent leakage of acoustic energy from the silicon into the oxide. Geometric softening is a technique that allows us to achieve simultaneous guiding of light and sound in a material system without under-etching and regardless of the bulk wave velocities. Although phonons and photons behave similarly in bulk media, their interactions with boundaries are markedly different. In particular, a solid-vacuum boundary geometrically softens the structural response of the material below and thus lowers the effective mechanical phase velocity [Fig. \ref{fig:guidingphonons}b]. This is the principle underpinning the 1D confinement of Rayleigh SAWs \cite{Auld1973,Oliner1973}. This mechanism was used in the 1970s in the megahertz range~\cite{Oliner1973,Lagasse,Ash1969} to achieve 2D confinement and was recently rediscovered for gigahertz phonons where it was found that both light and motion can be guided in unreleased silicon-on-insulator structures~\cite{Sarabalis2016}. More recently, fully 3D-confined acoustic waves have been demonstrated~\cite{Sarabalis2017a} with this approach on silicon-on-insulator where a narrow silicon fin, clamped to a silicon dioxide substrate, supports both localized photons and phonons.

\subsection{Phononic bandgaps}
\label{subsec:bandgap}

Structures patterned periodically, such as a silicon slab with a grid of holes, with a period $a$ close to half the phonons' wavelength $\Lambda = 2\pi/K$ result in strong mechanical reflections as in the optical case. At this $X$-point -- where $K = \pi/a$ -- in the dispersion diagram forward- and backward-traveling phonons are strongly coupled, resulting in the formation of a phononic bandgap [Fig. \ref{fig:guidingphonons}d] whose size scales with the strength of the periodic perturbation. The states just below and above the bandgap can be tuned by locally and smoothly modifying geometric properties of the lattice, resulting in the formation of line- or point-defects. This technique is pervasive in photonic crystals \cite{Joannopoulos2008} and was adapted to the mechanical case in the last decade \cite{Gorishnyy2005,Laude2005,Maldovan2006,Eichenfield2009,Safavi-Naeini2010g,Maldovan2013}. This led to the demonstration of optomechanical crystals that 3D-confine both photons and gigahertz phonons to a wavelength-scale suspended silicon nanobeams \cite{Eichenfield2009b,Chan2011b,Safavi-naeini2014}. In these experiments, confinement in one or two dimensions was obtained by periodic patterning of a bandgap structure, while in the remaining dimension confinement is due to the material being removed to obtain a suspended beam or film.

Conflicting demands similar to those discussed in section~\ref{sec:guideconfine}\ref{subsec:impedancemismatch} complicate the design of \emph{simultaneous} photonic-phononic bandgap structures \cite{Safavi-Naeini2010g}. For example, a hexagonal lattice of circular holes in a silicon slab as is often used in photonic bandgap cavities and waveguides, does not lead to a full phononic bandgap. Conversely, a rectangular array of cross-shaped holes in a slab as has been used to demonstrate full phononic bandgaps in silicon and other materials, does not support a photonic bandgap. Nonetheless, both one-dimensional \cite{Chan2011b} and two-dimensional crystals \cite{Safavi-naeini2014} with simultaneous photonic and phononic gaps have been proposed and demonstrated in technologically relevant material systems.

Beyond enabling 3D-confined wavelength-scale phononic cavities, phononic bandgaps also support waveguides or wires, which are 2D-confined defect states. These have been realized in silicon slabs with a pattern of cross-shaped holes supporting a full phononic bandgap, with an incorporated line defect within the bandgap material~\cite{Balram2015f,Fang2016,Fang2017b,Patel2017e}. Robustness to scattering is particularly important to consider in such nano-confined guided wave structures, since as in photonics, intermodal scattering due to fabrication imperfections increases with decreasing cross-sectional area of the guided modes \cite{Melati2014}. Single-mode phononic wires are intrinsically more robust as they remove all intermodal scattering except backscattering. They have been demonstrated to allow robust and low-loss phonon propagation over millimeter length scales~\cite{Patel2017e}. Currently both multi- and single-mode phononic waveguides are actively being considered as a means of generating connectivity and functionality in chip-scale solid-state quantum emitter systems based on defects in diamond \cite{Lemonde2018PhononWaveguides,Kuzyk2018PhononicWaveguides}.

\subsection{Other confinement mechanisms}
\label{subsec:othermechanisms}
The above mechanisms for confinement cover many if not most current systems. However, there are alternative mechanisms for photonic and phononic confinement, including but not limited to: bound states in the continuum \cite{Hsu2013,Gomis-Bresco2017,Plotnik2011}, Anderson localization \cite{Wiersma1997,Segev2013} and topological edge states \cite{Mousavi2015,Hafezi2013}. We do not cover these approaches here.

\subsection{Material limits}
\label{subsec:materiallimits}
Phononic confinement, propagation losses and lifetimes are limited by various imperfections such as geometric disorder \cite{VanLaer2015c,Safavi-naeini2014,Wolff2016,Patel2017a,Patel2017e,Sarabalis2018}, thermo-elastic and Akhiezer damping \cite{Auld1973,Ghaffari2013a}, two-level systems \cite{Behunin2016,Ramos2013a,Cleland2003a} and clamping losses \cite{VanLaer2015,Zhang2014EliminatingInterference,Anetsberger2008c}. Losses in 2D-confined waveguides are typically quantified by a propagation length $L_{\text{m}} = \alpha^{-1}_{\text{m}}$. In 3D-confined cavities one usually quotes linewidths $\gamma$ or quality factors $Q_{\text{m}}= \omegam/\gamma$. A cavity's internal loss rate can be computed from the decay length $L_{\text{m}}$ through $\gamma = \vm \alpha_{\text{m}}$ in high-finesse cavities with negligible bending losses \cite{VanLaer2016} with $\vm$ the mechanical group velocity. Mechanical propagation lengths in bulk crystalline silicon are limited to $L_{\text{m}} \approx 1 \, \text{cm}$ at room-temperature and at a frequency of $\omegam/(2\pi) = 1 \, \text{GHz}$ by thermo-elastic and Akhiezer damping. Equivalently, taking $\vm \approx 5000 \, \text{m/s}$ one can expect material-limited minimum linewidths of $\gamma/2\pi \approx 0.1 \, \text{MHz}$ and maximum quality factors of $Q_{\text{m}} \approx 10^{4}$ \cite{Auld1973,Ghaffari2013a} at $\omegam/(2\pi) = 1 \, \text{GHz}$.

These limits deteriorate rapidly at higher frequencies, typically scaling as $L_{\text{m}} \propto \omega^{-2}_{\text{m}}$ and $Q_{\text{m}} \propto \omega^{-1}_{\text{m}}$ \cite{Ghaffari2013a,Cleland2003a} or worse. This makes the $f_\text{m} \cdot Q_\text{m}$ product a natural figure of merit for mechanical systems. For gigahertz frequency resonators at room temperature, the highest demonstrated values of $f_\text{m} \cdot Q_\text{m}$  are on the order of $10^{13}$ in several materials \cite{Nguyen2007MEMSControl}. Intriguingly, the maximum length of time that a quantum state can persist inside a mechanical resonator with quality factor $Q_\text{m}$ at temperature $T$ is given by $t_\text{decoherence} =\frac{\hbar Q_\text{m}}{k T}$, and so requiring that the information survive for more than a mechanical cycle is equivalent to the condition $t_\text{decoherence}>\omega_\text{m}^{-1}$, or $f_\text{m} \cdot Q_\text{m} > 6\times 10^{12}~\text{Hz}$ at room temperature ~\cite{Aspelmeyer2014}. Recently, new loss mitigation mechanisms called ``strain engineering'' and ``soft clamping'' have been invented for megahertz mechanical resonators that enable mechanical quality factors and $f_\text{m} \cdot Q_\text{m}$ products beyond $10^{8}$ and $10^{15} \, \text{Hz}$ respectively under high vacuum but without refrigeration \cite{Norte2016MechanicalTemperature,Tsaturyan2017UltracoherentDilution,Ghadimi2018ElasticDissipation}. This unlocks exciting new possibilities for quantum-coherent operations at room temperature. Finally, we note that many material loss processes, with the exception of two-level systems \cite{Behunin2016}, vanish rapidly at low temperatures (section \ref{sec:stateoftheart}).

\begin{figure}[ht]
\includegraphics[width=\linewidth]{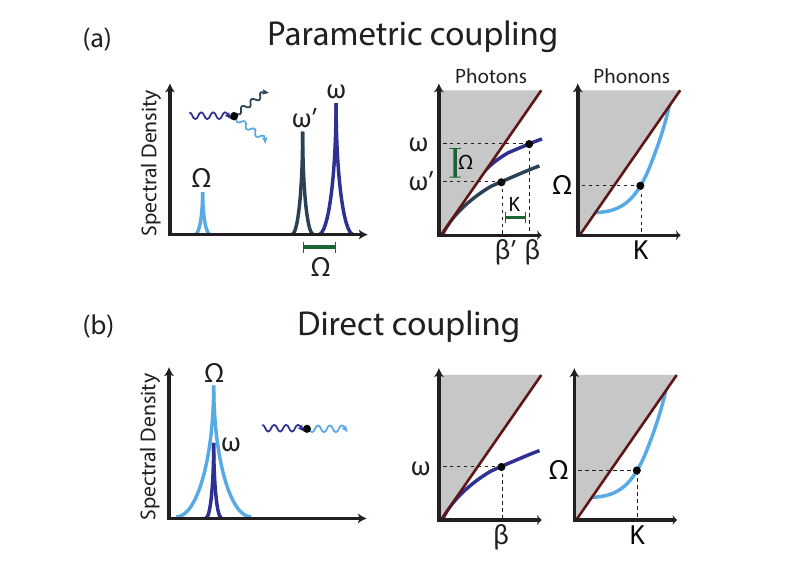}
\caption{\textbf{Generating and detecting phonons.} (a) Interactions between phonons and high-frequency photons occur through parametric three-wave mixing: two high-frequency photons couple to one phonon via third-order nonlinearities such as photoelasticity and the moving-boundary effect \cite{Rakich2012,Chan2012}. Interactions between low-frequency photons and a phonon can also occur through these mechanisms. Depending on which of the three waves is pumped, the interaction results in either down/up-conversion ($\delta a \, \delta b + \text{h.c.}$) or state-swapping ($\delta a \, \delta b^{\dagger} + \text{h.c.}$) events. The frequency-difference between the two high-frequency photons $\omega - \omega'$ must approximately equal the phononic frequency $\Omega$ for efficient parametric interactions to occur (left). In addition, in structures with translational symmetry the wavevector-difference between the two high-frequency photons $\beta - \beta'$ must also approximately equal the phononic wavevector $K$ for efficient coupling (right). Here we depict only DFD, SFD proceeds analogously but with minus signs replaced by plus signs. (b) Direct conversion via second-order nonlinearities such as piezoelectricity is possible when the photonic and phononic energies match. Stronger mechanical waves can typically be generated by direct conversion than by indirect three-wave mixing of two optical waves. A microwave photon can be converted into a phonon and subsequently into an optical photon by cascading two of these processes.}
\label{fig:generatingdetectingphonons}
\end{figure}

\section{Photon-phonon interactions}
\label{sec:photonphononinteractions}

In this section we describe the key mechanisms underpinning the coupling between photons and phonons. Photon-phonon interactions occur via two main mechanisms:
\begin{itemize}
\item Parametric coupling [Fig.\ref{fig:generatingdetectingphonons}a]: two photons and one phonon interact with each other in a three-wave mixing process as in Brillouin and Raman scattering and optomechanics, where the latter includes capacitive electromechanics.
\item Direct coupling [Fig.\ref{fig:generatingdetectingphonons}b]: one photon and one phonon interact with each other directly as in piezoelectrics.
\end{itemize}
The parametric three-wave mixing takes place via two routes:
\begin{itemize}
\item Difference-frequency driving (DFD): two photons with frequencies $\omega$ and $\omega'$ drive the mechanical system through a beat note at frequency $\omega - \omega' = \Omega \approx \omegam$ in the forces.
\item Sum-frequency driving (SFD): two photons with frequencies $\omega$ and $\omega'$ drive the mechanical system through a beat note at frequency $\omega + \omega'=\Omega \approx \omegam$ in the forces.
\end{itemize}
Three-wave DFD is the only possible mechanism when the photons and phonons have a large energy gap, as in interactions between phonons and optical photons. In contrast, microwave photons can interact with phonons through any of the three-wave and direct processes.

\subsection{Interactions between phonons and optical photons}
\label{subsec:3WMDFD}

Parametric DFD in a cavity is generally described by an interaction Hamiltonian of the form (see Appendix)
\begin{equation}
\label{eq:cavHamil}
\mathcal{H}_{\text{int}} = \hbar (\partialx \omegao) a^{\dagger}a x
\end{equation}
with $\partialx \omegao$ the sensitivity of the optical cavity frequency $\omegao$ to mechanical motion $x$ and $a$ the photonic annihilation operator. The terminology ``parametric'' refers to the parameter $\omegao$, essentially the photonic energy, being modulated by the mechanical motion \cite{Louisell1961QuantumI.,Bloembergen1964CouplingMedia,Shen1965,Yariv1965}, whereas the term ``three-wave mixing'' points out that there are three operators in the Hamiltonian given by equation \ref{eq:cavHamil}. This does not restrict the interaction to only three waves, as discussed further on. Describing the Hamiltonian $\mathcal{H}_{\text{int}}$ in this manner is a concise way of capturing all the consequences of the interaction between the electromagnetic field $a$ and the mechanical motion $x$. The detailed dynamics can be studied via the Heisenberg equations of motion defined by $\dot{a} = -\frac{i}{\hbar}[a,\mathcal{H}_{\text{int}}]$ when making use of the harmonic oscillator commutator $[a,a^{\dagger}]=1$ \cite{Aspelmeyer2014}. Since by definition $x = \xzp \left(\delta b + \delta b^{\dagger} \right)$ with $\xzp$ the mechanical zero-point fluctuations and $\delta b$ the phonon annihilation operator, this is equivalent to 
\begin{equation}
\label{eq:threewaveHamil}
\mathcal{H}_{\text{int}} = \hbar g_0 a^{\dagger}a (\delta b + \delta b^{\dagger})
\end{equation}
with
\begin{equation}
g_0 = (\partialx \omegao) \xzp
\end{equation}
the zero-point optomechanical coupling rate, which quantifies the shift in the optical cavity frequency $\omegao$ induced by the zero-point fluctuations $\xzp$ of the mechanical oscillator. Here we neglect the static mechanical motion \cite{Aspelmeyer2014,Wiederhecker2009,Rosenberg2009}. Achieving large $g_0$ thus generally requires small structures with large sensitivity $\partialx \omegao$ and zero-point motion $\xzp = (\hbar/(2 \omegam \meff))^{1/2}$, where  $\meff$ is the effective mass of the mechanical mode. This is brought about by ensuring a good overlap between the phononic field and the photonic forces acting on the mechanical system \cite{Chan2012,VanLaer2015,Rakich2012} and by focusing the photonic and phononic energy into a small volume to reduce $\meff$. There are typically separate bulk and boundary contributions to the overlap integral. The bulk contribution is associated with photoelasticity, while the boundary contribution results from deformation of the interfaces between materials \cite{Rakich2012,Qiu2013,Chan2012,Wolff2015}. Achieving strong interactions requires careful engineering of a constructive interference between these contributions \cite{VanLaer2015,Rakich2012,Qiu2013,Chan2012}. Optimized nanoscale silicon structures with mechanical modes at gigahertz frequencies typically have $\xzp \approx 1 \, \text{fm}$ and $g_0/(2\pi) \approx 1 \, \text{MHz}$ (section \ref{sec:stateoftheart}). The zero-point fluctuation amplitude increases with lower frequency leading to an increase in $g_0$: megahertz-frequency mechanical systems with  $g_0/(2\pi) \approx 10 \, \text{MHz}$ have been demonstrated \cite{Leijssen2015}.

\begin{figure*}[ht]
\centering
\includegraphics[width=\linewidth]{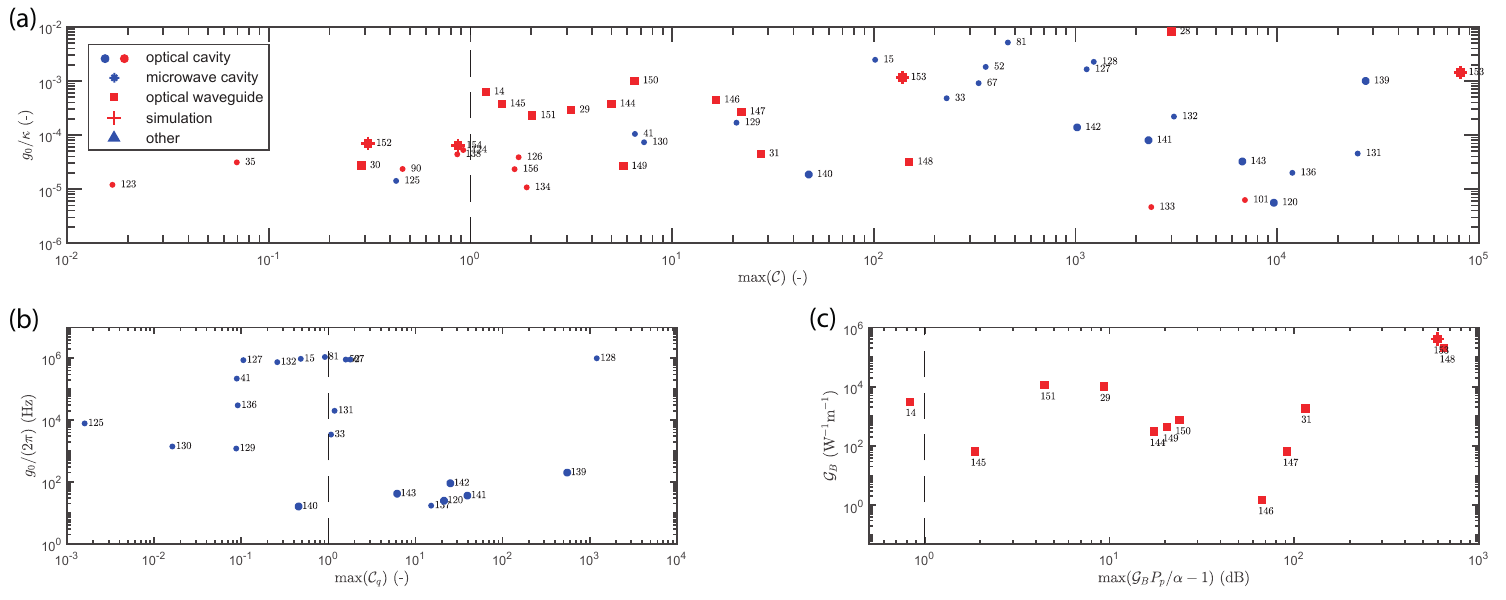}
\caption{\textbf{Interaction rates in 3D-confined cavities and 2D-confined waveguides.} (a) The dimensionless nonlinearity $g_0/\kappa$ versus the maximum cooperativity $\mathcal{C}$. The largest interaction rates are achieved in nanoscale cavities and waveguides. Cooperativities are typically highest in cold systems (blue) and can be as high in less tightly confined systems since they often have lower photonic and phononic decay rates than the smallest systems. (b) The zero-point coupling rate $g_0$ versus the maximum quantum cooperativity $\mathcal{C}_{q} = \mathcal{C}/\bar{n}_{\text{th}}$ with $\bar{n}_{\text{th}}$ the thermal phonon occupation for a selection of cold 3D-confined systems. It is significantly harder to reach $\mathcal{C}_{q}>1$ than $\mathcal{C}>1$. In particular, optical absorption easily heats up mechanical systems -- effectively increasing $\bar{n}_{\text{th}}$ \cite{Meenehan2015}. (c) The Brillouin gain coefficient $\mathcal{G}_{\text{B}}$ versus maximum net Brillouin gain $(\mathcal{G}_{\text{B}}P_{\text{p}}/\alpha - 1)$ for a selection of 2D-confined waveguides. Up to extrinsic cavity losses, $\mathcal{C} -1 =(\mathcal{G}_{\text{B}}P_{\text{p}}/\alpha - 1)$. Achieving this maximum Brillouin gain requires the waveguide to be have a length $L = 1/\alpha$ with $\alpha$ the optical propagation loss per meter. However, this is challenging as longer waveguides can suffer from an inhomogeneous broadening of the mechanical resonance due to atomic-scale disorder in the waveguide geometry.}
\label{fig:interactionrates}
\end{figure*}

The dynamics generated by the Hamiltonian of equation \ref{eq:threewaveHamil} can lead to a feedback loop. The beat note between two photons with slightly different frequencies $\omega$ and $\omega'$ generates a force that drives phonons at frequency $\omega - \omega' = \Omega$. Conversely, phonons modulate, at frequency $\Omega$, the optical field, scattering photons into up- and downconverted sidebands. This feedback loop can amplify light or sound, lead to electromagnetically induced transparency, or cooling of mechanical modes. In principle, this interaction can even cause strong nonlinear interactions at the few photon or phonon limit if $g_0/\kappa > 1$~\cite{Rabl2011PhotonSystems}, though current solid-state systems are more than two orders of magnitude away from this regime (see Figure~\ref{fig:interactionrates} and section \ref{sec:perspectives}).

Assuming $g_0/\kappa \ll 1$, valid in nearly all systems, we linearize the Hamiltonian of equation \ref{eq:threewaveHamil} by setting $a = \alpha + \delta a$ with $\alpha$ a classical, coherent pump amplitude, yielding
\begin{equation}
\label{eq:linearHamil}
\mathcal{H}_{\text{int}} = \hbar g\left(\delta a + \delta a^{\dagger} \right)\left(\delta b + \delta b^{\dagger} \right) 
\end{equation}
with $g = g_0 \alpha$ the enhanced interaction rate -- taking $\alpha$ real -- and $\delta a$ and $\delta b$ the annihilation operators representing photonic and phononic signals respectively. Often there are experimental conditions that suppress a subset of interactions present in Hamiltonian \ref{eq:linearHamil}. For instance, in sideband-resolved optomechanical cavities ($\omegam>\kappa$) a blue-detuned pump $\alpha$ sets up an entangling interaction
\begin{equation}
\label{eq:squeezingHamil}
\mathcal{H}_{\text{int}} = \hbar g \left(\delta a \delta b  + \delta a^{\dagger} \delta b^{\dagger}\right)
\end{equation}
that creates or annihilates photon-phonon pairs. Similarly, a red-detuned pump $\alpha$ sets up a beam-splitter interaction
\begin{equation}
\label{eq:beamsplitterHamil}
\mathcal{H}_{\text{int}} = \hbar g \left(\delta a \delta b^{\dagger} + \delta a^{\dagger} \delta b \right)
\end{equation}
that converts photons into phonons or vice versa. This beam-splitter Hamiltonian can also be realized by pumping the phononic instead of the photonic mode. In that case, $\alpha$ represents the phononic pump amplitude, whereas both $\delta a$ and $\delta b$ are then photonic signals.

In multi-mode systems, such as in 3D-confined cavities with several modes or in 2D-confined continuum systems, the interaction Hamiltonian is a summation or integration over each of the possible interactions between the individual photonic and phononic modes. For instance, linearized photon-phonon interactions in a 2D-confined waveguide with continuous translational symmetry are described by \cite{Sipe2016,Zoubi2016b,Rakich2018}:
\begin{equation}
\label{eq:linearHamilwg}
\mathcal{H}_{\text{int}} = \frac{\hbar}{\sqrt{2\pi}} \int \int \text{d}\beta\,\text{d}K \, \left(g_{\beta + K} a_{\beta} b_{K}  + g_{\beta - K} a_{\beta} b^{\dagger}_{K} + \text{h.c.}\right)
\end{equation}
In this case the three-wave mixing interaction rate $g_{\beta + K} = g_{0 | \beta + K} \alpha^{\star}_{\beta + K}$ is proportional to the amplitude $\alpha_{\beta + K}$ of the mode with wavevector $\beta + K$, which is usually considered to be pumped strongly. In contrast to the single-mode cavity described by equation \ref{eq:linearHamil}, in the waveguide case the symmetry between the two-mode-squeezing $\delta a \delta b$ and the beam-splitter $\delta a \, \delta b^{\dagger}$ terms is broken by momentum selection from the onset as generally $g_{\beta + K} \neq g_{\beta - K}$. The Hamiltonian of equation \ref{eq:linearHamilwg} assumes an infinitely long waveguide where phase-matching is strictly enforced. In contrast, a finite-length waveguide allows for interactions between a wider set of modes, although it suppresses those with a large phase-mismatch (see Appendix). In essence, shorter waveguides permit larger violations of momentum conservation. The momentum selectivity can enable non-reciprocal transport of both photons \cite{ZongfuYu2010,Kim2015,Sohn2018a,Kittlaus2018NonreciprocalModulation} and phonons \cite{Seif2018,Maldovan2013}. It is a continuum version of interference-based synthetic magnetism schemes using discrete optomechanical elements \cite{Peano2015,Fang2017b}.

Cavities can be realized by coiling up or terminating a 2D-confined waveguide with mirrors. Then the cavity's optomechanical coupling rate $g_0$ is connected to the waveguide's coupling rate $g_{0|\beta+K}$ by
\begin{equation}
\label{eq:linkg0}
g_0 = \frac{g_{0|\beta +K}}{\sqrt{L}}
\end{equation}
with $L$ the roundtrip length of the cavity (see Appendix). The parameters $g_{0|\beta +K}$ and $g_0$ are directly related to the so-called Brillouin gain coefficient $\mathcal{G}_{\text{B}}$ that is often used to quantify photon-phonon interactions in waveguides \cite{VanLaer2015,Kittlaus2016,Merklein2016}. In particular \cite{VanLaer2016},
\begin{equation}
\mathcal{G}_{\text{B}} = \frac{4g^{2}_{0|\beta +K}}{\vp \vs (\hbar \omega)\gamma}
\label{eq:linkcavwg}
\end{equation}
with $\vp$ and $\vs$ the group velocities of the interacting photons, $\hbar \omega$ the photon energy and $\gamma$ the phononic decay rate. Equations \ref{eq:linkg0} and \ref{eq:linkcavwg} enable comparison of the photon-phonon interaction strengths of waveguides and cavities. Since this gain coefficient depends on the mechanical quality factor $\Qm$ via $\gamma = \omegam/\Qm$, it is occasionally worth comparing waveguides in terms of the ratio $\mathcal{G}_{\text{B}}/\Qm$. The $g_0/(2\pi) \approx 1 \, \text{MHz}$ measured in silicon optomechanical crystals \cite{Safavi-Naeini2012} is via equation \ref{eq:linkcavwg} in correspondence with the $\mathcal{G}_{\text{B}}/\Qm \approx 10 \, \text{W}^{-1}\text{m}^{-1}$ measured in silicon nanowires at slightly higher frequencies \cite{VanLaer2015,VanLaer2015c}. Both $g_0$ and $\mathcal{G}_{\text{B}}/\Qm$ have an important dependence on mechanical frequency $\omegam$: lower-frequency structures are generally more flexible and thus generate larger interaction rates.

The Hamiltonians given in equations \ref{eq:squeezingHamil}, \ref{eq:beamsplitterHamil} and \ref{eq:linearHamilwg} describe a wide variety of effects. The detailed consequences of the three-wave mixing depend on the damping, intensity, dispersion and momentum of the interacting fields. Next, we describe some of the potential dynamics. We quantify the dissipation experienced by the photons and phonons with decay rates $\kappa$ and $\gamma$ respectively. The following regimes appear:
\begin{itemize}
\item Weak coupling: $g\ll \kappa+\gamma$. The phonons and photons can be seen as independent entities that interact weakly. A common figure of merit for the interaction is the cooperativity $\mathcal{C} = 4g^{2}/(\kappa \gamma)$, which quantifies the strength of the feedback loop discussed above. In particular, for $\mathcal{C}\gg 1$ the optomechanical back-action dominates the dynamics. The pair-generation Hamiltonian \ref{eq:squeezingHamil} generates amplification, whereas the beam-splitter interaction \ref{eq:beamsplitterHamil} generates cooling and loss. Whether the phonons or the photons dominantly experience this amplification and loss depends on the ratio $\kappa/\gamma$ of their decay rates. The linewidth of the phonons is effectively $(1\mp\mathcal{C})\gamma$ when $\kappa \ll \gamma$ where the minus-sign in $\mp$ holds for the amplification case (Hamiltonian \ref{eq:squeezingHamil}). In contrast, the linewidth of the photons is effectively $(1\mp\mathcal{C})\kappa$ when $\gamma \ll \kappa$. A lasing threshold is reached for the phonons or the photons when $\mathcal{C}=1$. In waveguide systems described by equation \ref{eq:linearHamilwg}, $\mathcal{C}=1$ is equivalent to the transparency point $\mathcal{G}_{\text{B}}P_{\text{p}}/\alpha = 1$ with $P_{\text{p}}$ the pump power and $\alpha$ the waveguide propagation loss per meter. In fact, interactions between photons and phonons in a waveguide can also be captured in terms of a cooperativity which is identical to $\mathcal{C}$ under only weakly restrictive conditions \cite{VanLaer2016}.

\item Strong coupling: $g \gg \kappa+\gamma$. The phonons and photons interact so strongly that they can no longer be considered independent entities. Instead, they form a photon-phonon polariton with an effective decay rate $(\kappa + \gamma)/2$. The beam-splitter interaction \ref{eq:beamsplitterHamil} sets up Rabi oscillations between photons and phonons with a period of $2\pi/g$ \cite{Groblacher2009,Verhagen2012,Aspelmeyer2014}. This is a necessary requirement for broadband intra-cavity state swapping, but is not strictly required for narrowband itinerant state conversion \cite{Wang2012,Wang2012b}.
\end{itemize}
Neglecting dynamics and when the detuning from the mechanical resonance is large ($\Delta \Omega \gg \gamma$), the phonon ladder operator is $\delta b = (g_0/\Delta \Omega) a^{\dagger}a$ such that Hamiltonian \ref{eq:threewaveHamil} generates an effective dispersive Kerr nonlinearity described by
\begin{equation}
\mathcal{H}_{\text{Kerr}} = \hbar \frac{g^{2}_0}{\Delta\Omega} a^{\dagger}a a^{\dagger}a
\end{equation}
This effective Kerr nonlinearity \cite{Ma2011,Pernice2009c,Aldana2013,Gea-Banacloche2010a,Butsch2012} is often much stronger than the intrinsic material nonlinearities. Thus a single optomechanical system can mediate efficient and tunable interactions between up to four photons in a four-wave mixing process that annihilates and creates two photons. The mechanics enhances the intrinsic optical material nonlinearities for applications such as wavelength conversion \cite{Hill2012} and photon-pair generation \cite{Riedinger2015}.

Additional dynamical effects exist in the multi-mode case. For instance, in a waveguide described by equation \ref{eq:linearHamilwg} there is a spatial variation of the photonic and phononic fields that is absent in the optomechanical systems described by equation \ref{eq:linearHamil}. This includes:
\begin{itemize}
\item The steady-state spatial Brillouin amplification of an optical sideband. This has been the topic of recent research in chip-scale photonic platforms. One can show that an optical Stokes sideband experiences a modified propagation loss $(1-\mathcal{C})\alpha$ with $\mathcal{C} = \mathcal{G}_{\text{B}}P_{\text{p}}/\alpha$ the waveguide's cooperativity \cite{VanLaer2016}. This Brillouin gain or loss is accompanied by slow or fast light \cite{Thevenaz2008}. Here we assumed an optical decay length exceeding the mechanical decay length, which is valid in nearly all systems. In the reverse case, the mechanical wave experiences a modified propagation loss $(1-\mathcal{C})\alpham$ and there is slow and fast sound \cite{VanLaer2016,Chen2016a,Otterstrom2018a}.
\item Traveling photonic pulses can be converted into traveling phononic pulses in a bandwidth far exceeding the mechanical linewidth. This is often called Brillouin light storage \cite{Zhu2007a,Merklein2017,Merklein2018,Stiller2018}. The traveling optical pump and signal pulses may counterpropagate or occupy different optical modes.
\end{itemize}
Several of these and other multi-mode effects have received little attention so far. This may change with the advent of new nanoscale systems realizing multi-mode and continuum Hamiltonians with strong coupling rates \cite{VanLaer2015,Kittlaus2016,Sipe2016,Zoubi2016b,Zoubi2017}.

\subsection{Interactions between phonons and microwave photons}
\label{subsec:2WM3WM}

The above section \ref{sec:photonphononinteractions}\ref{subsec:3WMDFD} on parametric three-wave DFD also applies to interactions between phonons and microwave photons. However, microwave photons may interact with phonons via two additional routes: (1) three-wave SFD and (2) direct coupling. In three-wave SFD, two microwave photons with a frequency below the phonon frequency $\omegam$ excite mechanical motion at the sum-frequency $\omega+\omega'=\Omega \approx \omegam$ \cite{Jones2013}. Such interactions can be realized in capacitive electromechanics, where the capacitance of an electrical circuit depends on mechanical motion. In particular, this sets up an interaction
\begin{equation}
\mathcal{H}_{\text{int}} = -\frac{(\partialx C) V^{2} x}{2}
\end{equation}
with $\partialx C$ the sensitivity of the capacitance $C(x)$ to the mechanical motion $x$ and $V$ the voltage across the capacitor. In terms of ladder operators we have $V = \Vzp (a + a^{\dagger})$ and $x = \xzp(\delta b + \delta b^{\dagger})$ such that
\begin{equation}
\mathcal{H}_{\text{int}} = \hbar g_0 \left(a^{\dagger}a + \frac{aa + a^{\dagger}a^{\dagger}}{2} + \frac{1}{2}\right)\left(\delta b + \delta b^{\dagger}\right)
\end{equation}
This interaction contains three-wave DFD (Hamiltonian \ref{eq:threewaveHamil}) as a subset via the $a^{\dagger}a$ term with an interaction rate $g_0$ given by
\begin{equation}
\hbar g_0 = -(\partialx C) \Vzp^{2} \xzp
\end{equation}
In addition to three-wave DFD, it also contains three-wave SFD via the $aa$ and $a^{\dagger}a^{\dagger}$ terms. These little explored terms enable electromechanical interactions beyond the canonical three-wave DFD optomechanical and Brillouin interactions.

Further, by applying a strong bias voltage $\Vb$ the capacitive interaction gets linearized: using $V = \Vb + \delta V$ and keeping only the $2\Vb \delta V$ term in $V^{2}$ yields
\begin{equation}
\mathcal{H}_{\text{int}} = - (\partialx C) \Vb \delta V x
\end{equation}
With $\delta V = \Vzp (\delta a + \delta a^{\dagger})$ this generates an interaction
\begin{equation}
\label{eq:linearHamilem}
\mathcal{H}_{\text{int}} = \hbar g (\delta a + \delta a^{\dagger}) (\delta b + \delta b^{\dagger})
\end{equation}
which is identical to the linearized optomechanics Hamiltonian in expression \ref{eq:linearHamil} with an interaction rate set by
\begin{equation}
\hbar g = -(\partialx C) \Vb \Vzp \xzp
\end{equation}
that is enhanced with respect to $g_0$ by $g = g_0 \alpha$ and $\alpha = \Vb/\Vzp$ the enhancement factor. The linearized Hamiltonian \ref{eq:linearHamilem} realizes a tunable, effective piezoelectric interaction that can directly convert microwave photons into phonons and vice versa. Piezoelectric structures are described by equation \ref{eq:linearHamilem} as well with an intrinsically fixed bias $\Vb$  determined by material properties.

Since $\Vzp = ((\hbar \omegamu)/(2C))^{1/2}$ with $\omegamu$ the microwave frequency and $C$ the total capacitance, the electromechanical coupling rate can be written as
\begin{equation}
g_0 = -\frac{\partialx C}{2 C} \omegamu \xzp
\end{equation}
or alternatively as $g_0 = (\partialx \omegamu) \xzp$ -- precisely as in section \ref{sec:photonphononinteractions}\ref{subsec:3WMDFD} but with the optical frequency $\omegao$ replaced by the microwave frequency $\omegamu$ with $\omegamu = 1/\sqrt{L_{\text{in}}C}$ and $L_{\text{in}}$ the circuit's inductance. Typically the capacitance $C = \Cm(x) + \Cp$ consists of a part that responds to mechanical motion $\Cm(x)$ and a part $\Cp$ that is fixed and usually considered parasitic. This leads to
\begin{equation}
g_0 = -\frac{\partialx \Cm}{2 \Cm} \etap \omegamu \xzp
\end{equation}
with $\etap = \Cm/C$ the participation ratio that measures the fraction of the capacitance responding to mechanical motion. For the canonical parallel-plate capacitor with electrode separation $s$, we have $\partialx \Cm = \Cm/s$ such that $g_0 = -\etap \xzp \omegamu/(2s)$. Similar to the optomechanics case, this often drives research towards small structures with large zero-point motion $\xzp$ and small electrode separation $s$. Contrary to the optomechanics case, however, increasing the participation ratio $\etap$ motivates increasing the size and thus the motional capacitance $\Cm$ of the structures until $\etap \approx 1$. In gigahertz-range microwave circuits with unity participation and electrode separations on the order of $s \approx 100 \, \text{nm}$, we have $g_0/(2\pi) \approx -10 \, \text{Hz}$, about a factor $\omegao/\omegamu \approx 10^{5}$ smaller than the optomechanical $g_0/(2\pi) \approx 1 \, \text{MHz}$ (section \ref{sec:photonphononinteractions}\ref{subsec:3WMDFD}). Despite the much smaller $g_0$, it is still possible to achieve large cooperativity $\mathcal{C} = 4g^{2}/(\kappa \gamma)$ in electromechanics as the typical microwave linewidths are much smaller and the enhancement factors $\alpha$ can be larger than in the optical case \cite{Fink2016a,Dieterle2016b,Kalaee2018QuantumCrystal}.

\begin{figure}[ht]
\centering
\includegraphics[width=\linewidth]{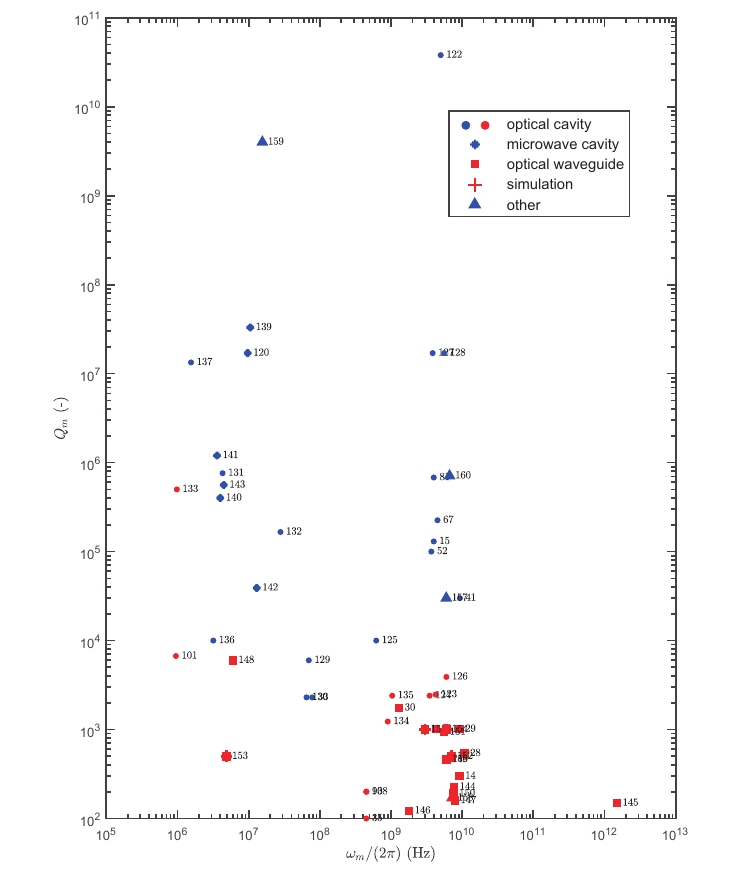}
\caption{\textbf{Mechanical quality factors.} The systems include opto- and electromechanical cavities and waveguides at room temperature (red), at a few kelvin (blue) and at millikelvin temperatures (blue). The highest quality factors are demonstrated in millikelvin cavities. The data points correspond to references \cite{Safavi-naeini2014,Safavi-Naeini2012,Chan2012,Chan2011b,Patel2017a,MacCabe2018,Bochmann2013c,Grutter2015,Liu2013b,Burek2016,Meenehan2014a,Meenehan2015,Verhagen2012,Riviere2011a,Schliesser2009a,Wilson2015a,Safavi-Naeini2013,Krause2015b,Sun2012d,Xiong2013,Fan2013,Purdy2012a,Groblacher2009,Sarabalis2017a,Leijssen2015,Williamson2016,Teufel2011,Weinstein2014a,Wollman2015a,Pirkkalainen2015a,Dieterle2016b,Fink2016b,VanLaer2015,VanLaer2015c,Shin2013b,Kittlaus2016,Pant2011,Rong2007,Kobyakov2009,Kang2009,Abedin2006,Butsch2014,Kittlaus2017,Morrison2017,VanLaer2017b,Mirnaziry2015,VanLaer2014,Wolff2014,VanLaer2017a,Otterstrom2018,Arrangoiz-Arriola2018a,VanLaer2018,Goryachev2012,Chu2017QuantumQubits}.}
\label{fig:mechanicalquality}
\end{figure}

\section{State of the art}
\label{sec:stateoftheart}
Here we give a concise overview of the current state of the art in opto- and electromechanical systems by summarizing the parameters obtained in about fifty opto- and electromechanical cavities and waveguides. First, we plot the mechanical quality factors as a function of mechanical frequency [Fig.\ref{fig:mechanicalquality}] including room temperature (red) and cold (blue) systems. As discussed in section \ref{sec:guideconfine}\ref{subsec:materiallimits}, cold systems usually reach much higher quality factors. The current record is held by a $5 \, \text{GHz}$ silicon optomechanical crystal with $\Qm > 10^{10}$, yielding a lifetime longer than a second \cite{MacCabe2018} at millikelvin temperatures. Measuring these quality factors requires careful optically pulsed read-out techniques, as the intrinsic dissipation of continuous-wave optical photons easily heats up the mechanics thus destroying its coherence \cite{Meenehan2015}. Comparably high quality factors are measured electrically in quartz and sapphire at lower frequencies \cite{Goryachev2012,Galliou2013}. It is an open question whether these extreme lifetimes have reached intrinsic material limits. The long lifetimes make mechanical systems attractive for delay lines and qubit storage \cite{Cleland2004b} (section \ref{sec:perspectives}).

Next, we look at the coupling strengths in these systems [Fig.\ref{fig:interactionrates}]. As discussed in section \ref{sec:photonphononinteractions}, a few different figures of merit are commonly used depending on the type of system. We believe the dimensionless ratios $g_0/\kappa$ and the cooperativity $\mathcal{C}$ are two of the most powerful figures of merit (section \ref{sec:perspectives}). The ratio $g_0/\kappa$ determines the single-photon nonlinearity, the energy-per-bit in optical modulators as well as the energy-per-qubit in microwave-to-optical photon converters. The cooperativity $\mathcal{C}$ must be unity for efficient state conversion as well as for phonon and photon lasing. In the context of waveguides it measures the maximum Brillouin gain as $\mathcal{C} = \mathcal{G}_{\text{B}} P_{\text{p}}/\alpha$ \cite{VanLaer2016}.

Thus we compute $g_0/\kappa$ for about fifty opto- and electromechanical cavities and waveguides [Fig.\ref{fig:interactionrates}a]. We convert the waveguide Brillouin coefficients $\mathcal{G}_{\text{B}}$ to $g_0$ via expressions \ref{eq:linkg0} and \ref{eq:linkcavwg} by estimating the minimum roundtrip length $L$ a cavity made from the waveguide would have. In addition, we convert the waveguide propagation loss $\alpha$ to the intrinsic loss rate $\kappain = \alpha \vg$ with $\vg$ the group velocity. This brings a diverse set of systems together in single figure. No systems exceed $g_0/\kappa\approx 0.01$, with the highest values obtained in silicon optomechanical crystals \cite{Safavi-Naeini2012,Chan2011b}, Brillouin-active waveguides \cite{VanLaer2015,VanLaer2015c,Kittlaus2016b} and Raman cavities \cite{Rong2007}. There is no strong relation between $g_0/\kappa$ and $\mathcal{C}$: systems with low interactions rates $g_0$ often have low decay rates $\kappa$ and $\gamma$ as well since they do not have quite as stringent fabrication requirements on the surface quality.

The absolute zero-point coupling rates $g_0$ illustrate the power of moving to the nanoscale. We plot them as a function of the maximum quantum cooperativity $\mathcal{C}_{\text{q}} = \mathcal{C}/\bar{n}_{\text{th}}$ with $\bar{n}_{\text{th}}$ the thermal phonon occupation [Fig.\ref{fig:interactionrates}b]. When $\mathcal{C}_{\text{q}}>1$, the state transfer between photons and phonons takes place more rapidly than the mechanical thermal decoherence \cite{Aspelmeyer2014}. This is a requirement for hybrid quantum systems such as efficient microwave-to-optical photon converters (section \ref{sec:perspectives}). There are several chip-scale electro- and optomechanical systems that obtained $\mathcal{C}_{\text{q}}>1$, with promising values demonstrated in silicon photonic crystals. A main impediment to large quantum cooperativities in optomechanics is the heating of the mechanics caused by optical absorption \cite{Meenehan2015}.

Further, we give an overview of the Brillouin coefficients $\mathcal{G}_{\text{B}}$ found in 2D-confined waveguides [Fig.\ref{fig:interactionrates}c]. The current record $\mathcal{G}_{\text{B}} = 10^{4} \, \text{W}^{-1}\text{m}^{-1}$ in the gigahertz range was measured in a suspended series of silicon nanowires \cite{VanLaer2015c}. However, larger Brillouin amplification was obtained with silicon and chalcogenide rib waveguides which have disproportionately lower optical propagation losses $\alpha$ and can handle larger optical pump powers $P_{\text{p}}$ \cite{Kittlaus2016b,Pant}. We stress that the maximum Brillouin gain is identical to the cooperativity \cite{VanLaer2016}. They are both limited by the maximum power and electromagnetic energy density the system in question can withstand. At room temperature in silicon, the upper limit is usually set by two-photon and free-carrier absorption \cite{VanLaer2015,Kittlaus2016,Wolff2015}. Moving beyond the two-photon bandgap of 2200 nm in silicon or switching to materials such as silicon nitride, lithium niobate or chalcogenides can drastically improve the power handling \cite{Wolff2015,Kuyken2011,Kuyken2016,Pant2011}. In cold systems, it is instead set by the cooling power of the refrigerator and the heating of the mechanical system \cite{Meenehan2015}. Another challenge for 2D-confined waveguides is the inhomogeneous broadening of the mechanical resonance. This arises from atomic-scale fluctuations in the waveguide geometry along its length effectively smearing out the mechanical response \cite{Wolff2016,VanLaer2015,VanLaer2015c,Kittlaus2016}. Finally, compared to gigahertz systems, flexible megahertz mechanical systems give much higher efficiencies of $\mathcal{G}_{\text{B}} \approx 10^{6} \, \text{W}^{-1}\text{m}^{-1}$ as measured in dual-nanoweb \cite{Koehler2017} fibers and of $\mathcal{G}_{\text{B}}\approx 10^{9} \, \text{W}^{-1}\text{m}^{-1}$ as predicted in silicon double-slot waveguides \cite{VanLaer2017}.

\section{Perspectives}
\label{sec:perspectives}
\subsection{Single-photon nonlinear optics}
\label{subsec:singlephotonNLO}
The three-wave mixing interactions discussed in section \ref{sec:photonphononinteractions} in principle enable single-photon nonlinear optics in opto- and electromechanical systems~\cite{Rabl2011PhotonSystems,Nunnenkamp2011Single-PhotonOptomechanics,Komar2013Single-photonOptomechanicsb}. For instance, in the photon blockade effect a single incoming photon excites the motion of a mechanical system in a cavity, which then shifts the cavity resonance and thus blocks the entrance of another photon. Realizing such quantum nonlinearities sets stringent requirements on the interaction strengths and decay rates.

For instance, in an optomechanical cavity the force exerted by a single photon is $\langle F \rangle = -\langle \partialx \mathcal{H}_{\text{int}}\rangle = -\hbar (g_0/\xzp)\langle a^{\dagger}a \rangle = -\hbar (g_0/\xzp)$. To greatly affect the optical response seen by another photon impinging on the cavity, this force must drive a mechanical displacement that shifts the optical resonance by about a linewidth $\kappa$ or  $\xpi = \kappa/(\partialx \omegao) = (\kappa/g_0)\xzp$. In other words, we require $F/(\meff \omegam^{2}) = \xpi$ which leads to $\vartheta_{\text{cav}} \equiv 4 g_0^{2}/(\kappa \omegam) \approx \pi$ where $\vartheta_{\text{cav}}$ is the mechanically-mediated cross-phase shift experienced by the other photon assuming critical coupling to the cavity. This extremely challenging condition is relaxed when two photonic modes with a frequency difference $\Delta\omega$ roughly resonant with the mechanical frequency are used. In this case, the mechanical frequency can be replaced by the detuning from the mechanical resonance in the above expressions: $\omegam \rightarrow 2\Delta\Omega$ with the detuning $\Delta\Omega = \Delta\omega - \omegam$. This enhances the shift per photon so that quantum nonlinearities are realized at \cite{Ludwig2012d,Komar2013Single-photonOptomechanicsb}
\begin{equation}
\vartheta_{\text{cav}} = \frac{2g_0^{2}}{\kappa \Delta \Omega} \approx \pi
\end{equation}
with $\Delta \Omega \ll \omegam$. The photon blockade effect also requires sideband-resolution ($\Delta\Omega > \kappa$) so
\begin{equation}
\frac{g_0}{\kappa} > 1
\end{equation}
is generally a necessary condition for single-photon nonlinear optics with opto- and electromechanical cavities \cite{Aspelmeyer2014,Kronwald2013}. In the case of 2D-confined waveguides, it can similarly be shown \cite{Zoubi2017} that a single photon drives a mechanically-mediated cross-Kerr phase shift
\begin{equation}
\label{eq:wgcrossKerr}
\vartheta_{\text{wg}} = \frac{g^{2}_{0|\beta +K}}{\vg \Delta \Omega}
\end{equation}
on another photon with $\vg$ the optical group velocity (see Appendix). The cross-Kerr phase-shift $\vartheta_{\text{wg}}$ can be enhanced drastically by reducing the group velocity $\vg$ via Brillouin slow light \cite{Zoubi2017,Thevenaz2008,Okawachi2005}. If sufficiently large, the phase-shifts $\vartheta_{\text{cav}}$ and $\vartheta_{\text{wg}}$ can be used to realize controlled-phase gates between photonic qubits -- an elementary building block for quantum information processors \cite{Brod2016,Brod2016a,Gea-Banacloche2014a,Zoubi2017}. Using equation \ref{eq:linkg0}, we have 
\begin{equation}
\label{eq:finnesecavwg}
\vartheta_{\text{cav}} = \frac{\mathcal{F}}{\pi} \vartheta_{\text{wg}}
\end{equation}
with $\mathcal{F} = 2\pi/(\kappa T_{\text{rt}})$ the cavity finesse and $T_{\text{rt}}$ the cavity roundtrip time. Therefore, cavities generally yield larger single-photon cross-Kerr phase shifts than their corresponding optomechanical waveguides.

Currently state-of-the-art solid-state and sideband-resolved ($\omega_\text{m}>\kappa$) opto- and electromechanical systems yield at best $g_0/\kappa \approx 0.01$ in any material [Fig.\ref{fig:interactionrates}]. Significant advances in $g_0$ may be made in e.g. nanoscale slotted structures \cite{Leijssen2015,Schneider2016StrongRatio,VanLaer2014}, but it remains an open challenge to not only increase $g_0$ but also $g_0/\kappa$ by a few orders of magnitude \cite{Baker2014PhotoelasticResonators}. Beyond exploring novel structures, other potential approaches include effectively boosting $g_0/\kappa$ by parametrically amplifying the mechanical motion \cite{Lemonde2016}, by employing delayed quantum feedback \cite{Wang2017a} or via collectively enhanced interactions in optomechanical arrays \cite{Xuereb2012,Gartner2018IntegratedMembranes}. Although single-photon nonlinear optics may be out of reach for now, many-photon nonlinear optics can be enhanced very effectively with mechanics. Specifically, mechanics realizes Kerr nonlinearities orders of magnitude beyond those of typical intrinsic material effects. This is especially so for highly flexible, low-frequency mechanical systems \cite{Roels2009a,VanThourhout2010,Pernice2009c,Ma2011} but has been shown in gigahertz silicon optomechanical cavities and waveguides as well \cite{Hill2012,VanLaer2015}.

\subsection{Efficient optical modulation}
\label{subsec:EOM}
Phonons provide a natural means for the spatiotemporal modulation of optical photons via electro- and optomechanical interactions. Hybrid circuits that marry photonic and phononic excitations give us access to novel opto-electro-mechanical systems. Two aspects of the physics make phononic circuits very attractive for the modulation of optical fields.

First, there is excellent spatial matching between light and sound. As touched upon above, the wavelengths of microwave phonons and telecom photons are both about a micron in technologically relevant materials such as silicon. The matching follows from the four to five orders of magnitude difference between the speed of sound and the speed of light. Momentum conservation, i.e. phase-matching, between phonons and optical photons (as discussed in section \ref{sec:photonphononinteractions}) is key for non-reciprocal nonlinear processes and modulation schemes with traveling phonons \cite{Poulton2012,ZongfuYu2010,Sohn2018a,Fang2017a}. 

Second, the optomechanical nonlinearity is strong and essentially lossless. Small deformations can induce major changes on the optical response of a system. For instance, in an optomechanical cavity (equation \ref{eq:cavHamil}) the mechanical motion required to encode a bit onto a light field has an amplitude of approximately $\xpi=(\kappa/g_0)\xzp$. Generating this motion requires energy, and this corresponds to an energy-per-bit $E_{\text{bit}} = \meff \omegam^{2}\xpi^{2}/2$ which we rewrite as
\begin{equation}
\label{eq:energyperbit}
E_{\text{bit}} = \frac{\hbar \omegam}{4} \left(\frac{\kappa}{g_0}\right)^{2}
\end{equation}
Thus the energy-per-bit also depends on the dimensionless quantity $g_0/\kappa$: a single phonon can switch a photon when this quantity reaches unity, in agreement with section \ref{sec:perspectives}\ref{subsec:singlephotonNLO}. For silicon optomechanical crystals with $g_0/\kappa \approx 10^{-3}$, this yields $E_{\text{bit}} \approx 1 \, \text{aJ/bit}$: orders of magnitude more efficient than commonly deployed electro-optic technologies \cite{Miller2017b}.

The similarity between the fundamental interactions in optomechanics \cite{Aspelmeyer2014} and electro-optics \cite{Tsang2010a,Tsang2011a} allows to compare the two types of modulation head-to-head. In particular, in an optical cavity made of an electro-optic material the voltage drop across the electrodes required to encode a bit is $\Vpi = \kappa/(\partialV \omegao) =  (\kappa/g_0)\Vzp$ with $g_0 = (\partialV \omegao) \Vzp$ the electro-optic interaction rate \cite{Tsang2010a,Tsang2011a}, which is defined analogously to the optomechanical interaction rate. It is the parameter appearing in the interaction Hamiltonian $\mathcal{H}_{\text{int}} = \hbar g_0 a^\dagger a(b+b^\dagger)$, with $b+b^\dagger$ now proportional to the voltage across the capacitor of a microwave cavity \cite{Tsang2010a,Tsang2011a}. The required $\Vpi$ corresponds to an energy-per-bit $E_{\text{bit}} = C \Vpi^{2}/2$ which again can be rewritten as expression \ref{eq:energyperbit}. Electro-optic materials such as lithium niobate \cite{Wang2018NanophotonicModulators} may yield up to $g_0/(2\pi) \approx 10 \, \text{kHz}$, corresponding to an energy-per-bit $E_{\text{bit}} \approx 10 \, \text{fJ/bit}$ keeping the optical linewidth $\kappa$ constant -- on the order of today's world records \cite{Miller2017b}.

Although full system demonstrations using mechanics for electro-optic modulation are lacking, based on estimates like these we believe that mechanics will unlock highly efficient electro-optic systems. The expected much lower energy-per-bit implies that future electro-opto-mechanical modulators could achieve much higher bitrates at fixed power, or alternatively, much lower dissipated power at fixed bitrate than current direct electro-optic modulators. Although the mechanical linewidth does not enter expression \ref{eq:energyperbit}, bandwidths of a single device are usually limited by the phononic quality factor or transit time across the device. Interestingly, the mechanical displacement corresponding to the estimated $1 \, \text{aJ/bit}$ is only $\xpi \approx 10 \, \text{pm}$.

Here we highlighted the potential for optical modulation based on mechanical motion at gigahertz-frequencies. However, similar arguments can be made for optical switching networks based on lower frequency mechanical structures. In particular, voltage-driven capacitive or piezoelectric optical phase-shifters exploiting mechanical motion do not draw static power and can generate large optical phase shifts in small devices \cite{VanAcoleyen2012,Pitanti2015StrongCavity,Winger2011APlatform,Poot2014BroadbandChip,Jin2018PhaseContraction,Errando-Herranz2015Low-powerFilter,Midolo2018a,Seok2016Large-scaleCouplers}. These ``photonic MEMS'' are thus an attractive elementary building block in reconfigurable and densely integrated photonic networks used for high-dimensional classical \cite{Miller2013Self-aligningCoupler.,Miller2015PerfectComponents,Miller2017b,Carolan2015UniversalOptics,Shen2017DeepCircuits} and quantum \cite{Rudolph2017WhyComputing,Shadbolt2012GeneratingCircuit,Qiang2018Large-scaleProcessing} photonic information processors. They may meet the challenging power- and space-constraints involved in running a complex programmable network.

Demonstrating fully integrated acousto-optic systems requires that we properly confine, excite and route phonons on a chip. Among the currently proposed and demonstrated systems are acousto-optic modulators \cite{Sohn2018a} as well as optomechanical beam-steering systems \cite{Sarabalis2018,Errando-Herranz2018Low-powerGratings}. Besides showing the power of sound to process light with minuscule amounts of energy, these phononic systems have features that are absent in competing approaches. For instance, gigahertz traveling mechanical waves with large momentum naturally enable non-reciprocal features in both modulators \cite{Sohn2018a} and beam-steering systems \cite{Sarabalis2018}. This is essential for isolators and circulators based on indirect photonic transitions \cite{Jalas2013,Sounas2017Non-reciprocalModulation,Miri2017OpticalCoupling,Kim2015,Verhagen2017OptomechanicalNonreciprocity,Kittlaus2018NonreciprocalModulation}.

In order to realize these and other acousto-optic systems, it is crucial to efficiently excite mechanical excitations on the surface of a chip. In this context, electrical excitation is especially promising as it allows for stronger mechanical waves than optical excitation. With optical excitation of mechanical waves, the flux of phonons is upper-bounded by the flux of optical photons injected into the structure. The ratio of photon to phonon energy limits the mechanical power to less than a microwatt, corresponding to 10-100 milliwatts of optical power. Nevertheless, proof-of-concept demonstrations~\cite{Kim2015,Kittlaus2018NonreciprocalModulation} have successfully generated non-reciprocity on a chip using optically generated phonons. In contrast, microwave photons have a factor $10^{5}$ larger fluxes than optical photons for the same power. Therefore, microwave photons can drive milliwatt-level mechanical waves in nanoscale cavities and waveguides. Such mechanical waves can have displacements up to a nanometer and strains of a few percent -- close to material yield strengths.

Electrical generation of gigahertz phonons in nanoscale structures has received little attention so far, especially in non-piezoelectric materials such as silicon and silicon nitride. As discussed in section \ref{sec:photonphononinteractions}\ref{subsec:2WM3WM}, this can be realized either via capacitive or via piezoelectric electromechanics. Capacitive approaches work in any material \cite{Unterreithmeier2009} and have recently been demonstrated in a silicon photonic waveguide~\cite{VanLaer2018}. They require small capacitor gaps and large bias voltages to generate effects of magnitude comparable to piezoelectric approaches. More commonly, piezoelectrics such as lithium niobate, aluminum nitride \cite{Tadesse2014b,Xiong2013a} and lead-zirconate titanate can be used as the photonic platform, or be integrated with existing photonic platforms such as silicon and silicon nitride in order to combine the best of both worlds \cite{Alexander2018a,Weigel2018,Sarabalis2018,Siddiqui2018LambFilms}. Such hybrid integration typically comes with challenging incompatibilities in material properties \cite{Marshall2018HeterogeneousPhotonics}, especially when more than one material needs to be integrated on a single chip. Efficient electrically-driven acoustic waves in photonic structures have the potential to enable isolation and circulation with an optical bandwidth beyond $1 \, \text{THz}$ -- limited only by optical walk-off \cite{Yu2009OpticalTransitions,ZongfuYu2010,Fang2012,Shi2017OpticalModes,Kittlaus2018NonreciprocalModulation}.

\subsection{Hybrid quantum systems}
\label{subsec:QEOM}
Strain and displacement alter the properties of many different systems and therefore provide excellent opportunities for connecting dissimilar degrees of freedom. In addition, mechanical systems can possess very long coherence times and can be used to store quantum information. In the field of hybrid quantum systems, researchers find ways to couple different degrees of freedom over which quantum control is possible to scale up and extend the power of quantum systems. Realizing hybrid systems by combining mechanical elements with other excitations is a widely pursued research goal. Studies on both static tuning of quantum systems using nanomechanical forces~\cite{Lee2017TopicalDiamond,Barson2017NanomechanicalDiamond,Sohn2018ControllingEnvironment} as well as on quantum dynamics mediated by mechanical resonances and waveguides  \cite{OConnell2010,Lee2017TopicalDiamond,Andrews2014c,Lemonde2018PhononWaveguides,Pechal2018SuperconductingStorage} are being pursued.

Among the emerging hybrid quantum systems, microwave-to-optical photon converters utilizing mechanical degrees of freedom have attracted particular interest recently \cite{Safavi-Naeini2011,Bochmann2013c,Bagci2014,Andrews2014c,Rueda2016b,Vainsencher2016c}. In particular, one of the leading platforms to realize scalable, error-corrected quantum processors \cite{Preskill2012a,Preskill2018b} are superconducting microwave circuits in which qubits are realized using Josephson junctions \cite{Devoret2013SuperconductingOutlook,Barends2014a} in a platform compatible with silicon photonics~\cite{Keller2017}. To suppress decoherence, these microwave circuits are operated at millikelvin temperatures inside dilution refrigerators. Heat generation must be restricted in these cold environments \cite{Krinner2018}. The most advanced prototypes currently consist of on the order of fifty qubits on which gates with at best $0.1\%$ error rates can be applied \cite{Krinner2018,Neill2018}. Scaling up these systems to millions of qubits, as required for a fully error-corrected quantum computer, is a formidable unresolved challenge \cite{Preskill2018b}. Also, the flow of microwave quantum information is hindered outside of the dilution refrigerators by the microwave thermal noise present at room temperature \cite{Vermersch2017a,Xiang2017}. Optical photons travel for kilometers at room temperature along today's optical fiber networks. Thus quantum interfaces that convert microwave to optical photons with high efficiency and low noise should help address the scaling and communication barriers hindering microwave quantum processors. They may pave the way for distributed quantum computing systems or a ``quantum internet'' \cite{Kimble2008}. Besides, such interfaces would give optical systems access to the large nonlinearities generated by Josephson junctions, which enables a new approach for nonlinear optics.

The envisioned microwave-to-optical photon converters are in essence electro-optic modulators that operate on single photons and preserve entanglement \cite{Tsang2010a}. They exploit the beam-splitter Hamiltonian discussed in section \ref{sec:photonphononinteractions} to swap quantum states from the microwave to the optical domain and vice versa. To realize a microwave-to-optical photon converter, one can start from a classical electro-optic modulator and modify it to protect quantum coherence. Several proposals aim to achieve this by coupling a superconducting microwave cavity to an optical cavity made of an electro-optic material. For instance, the beam-splitter Hamiltonian can be engineered by injecting a strong optical pump red-detuned from the cavity resonance in an electro-optic cavity. In order to suppress undesired Stokes scattering events, the frequency of the microwave cavity needs to exceed the optical cavity linewidth, i.e. sideband-resolution is necessary. In this scenario, continuous-wave state conversion with high fidelity requires an electro-optic cooperativity $\mathcal{C}_{\text{eo}}$ close to unity:
\begin{equation}
\label{eq:coopunity}
\mathcal{C}_{\text{eo}} = \frac{4g^{2}_{0} |\alpha|^{2}}{\kappa \gamma_{\mu}} = 1
\end{equation}
with $g_0$ the electro-optic interaction rate as defined in the previous section, $|\alpha|^{2}$ the number of optical pump photons in the cavity and $\gamma_{\mu}$ the microwave cavity linewidth. The quantum conversion is accompanied by an optical power dissipation $P_{\text{diss}} = \hbar \omegao |\alpha|^{2} \kappain$ with $\kappain$ the intrinsic decay rate of the optical cavity. Operating the converter in a bandwidth of $\gamma_{\mu}$ and inserting condition \ref{eq:coopunity}, this leads to an energy-per-qubit of
\begin{equation}
\label{eq:energyperqubit}
E_{\text{qbit}} = \frac{\hbar \omegao}{4} \left(\frac{\kappa \kappain}{g^{2}_0} \right)
\end{equation}
which is the quantum version of the energy-per-bit \ref{eq:energyperbit}. This yields an interesting relation between the efficiency of classical and quantum modulators
\begin{equation}
\frac{E_{\text{qbit}}}{E_{\text{bit}}} \approx \frac{\omegao}{\omegam}
\end{equation}
We stress that $E_{\text{qbit}}$ is the optical dissipated energy in a quantum converter, whereas $E_{\text{bit}}$ is the microwave or mechanical energy necessary to switch an optical field in a classical modulator \cite{Pechal2017a}. The quantum electro-optic modulator dissipates roughly five orders of magnitude more energy per converted qubit as it requires an optical pump field to drive the conversion process. Strategies developed to minimize $E_{\text{bit}}$, as pursued for decades by academic groups and the optical communications industry, also tend to minimize $E_{\text{qbit}}$. Recently a coupling rate of $g_{0}/(2\pi) = 310 \, \text{Hz}$ was demonstrated in an integrated aluminum nitride electro-optic resonator \cite{Fan2018a}. Switching to lithium niobate and harnessing improvements in the electro-optic modal overlap may increase this to $g_{0}/(2\pi) \approx 10 \, \text{kHz}$, corresponding to $E_{\text{qbit}} \approx 1 \, \text{nJ/qbit}$. Electro-optic polymers \cite{Kieninger2018} may yield higher interaction rates $g_{0}$ but bring along challenges in optical and microwave losses $\kappa$ and $\gamma_{\mu}$. Cooling powers of roughly $10 \, \mu\text{W}$ at the low temperature stage of current dilution refrigerators \cite{Krinner2018} imply that conversion rates with common electro-optic materials will likely not exceed about $10 \, \text{kqbits/s}$.

Considering that the $g_0/\kappa$ demonstrated optomechanical devices is much larger than those found in electro-optic systems, and following a reasoning similar to that presented in section \ref{sec:perspectives}\ref{subsec:EOM} for classical modulators, it is likely that microwave-to-optical photon converters based on mechanical elements as intermediaries will be able to achieve large efficiencies. It has been theoretically shown that electro-opto-mechanical cavities with dynamics described in section \ref{sec:photonphononinteractions} allow for efficient state transduction between microwave and optical fields when
\begin{equation}
\label{eq:mechconverter}
\mathcal{C}_{\text{em}} \approx \mathcal{C}_{\text{om}} \gg 1
\end{equation}
with $\mathcal{C}_{\text{em}}$ and $\mathcal{C}_{\text{om}}$ the electro- and optomechanical cooperativities. Noiseless conversion additionally requires negligible thermal microwave and mechanical occupations \cite{Wang2012,Wang2012b,Safavi-Naeini2011}. Since the dominant dissipation still arises from the optical pump, the energy-per-qubit can still be expressed as in equation \ref{eq:energyperqubit} for a electro-opto-mechanical cavity. Given the large nonlinearity $g_0/\kappa$ enabled by nanoscale mechanical systems [Fig.\ref{fig:interactionrates}], we expect conversion rates up to $100 \, \text{Mqbits/s}$ are feasible by operating multiple electro-opto-mechanical photon converters in parallel inside the refrigerator. State-of-the-art integrated electro- and opto-mechanical cavities have achieved $\mathcal{C}_{\text{em}} > 1$ and $\mathcal{C}_{\text{om}}>1$ in separate systems [Fig.\ref{fig:interactionrates}]. It is an open challenge to achieve condition \ref{eq:mechconverter} in a single integrated electro-opto-mechanical device.

Finally, the long lifetimes and compact nature of mechanical systems also makes them attractive for the storage of classical and quantum information \cite{Cleland2004b,Arrangoiz-Arriola2016a,Chu2017QuantumQubits,Gustafsson2014c,Ekstrom2017,Arrangoiz-Arriola2018a,Ekinci2005NanoelectromechanicalSystems,Pechal2018SuperconductingStorage,Mahboob2008}. Mechanical memories are currently pursued both with purely electromechanical \cite{OConnell2010,Chu2017QuantumQubits,Arrangoiz-Arriola2018a} and purely optomechanical \cite{Zhu2007a,Merklein2017} systems. Interfaces between mechanical systems and superconducting qubits may lead to the generation of non-classical states of mesoscopic mechanical systems \cite{Chu2018a,Aspelmeyer2014,Davis2018PaintingPhotons}, probing the boundary between quantum and classical behavior.

\subsection{Microwave signal processing}
In particular in the context of wireless communications, compact and cost-effective solutions for radio-frequency (RF) signal processing are rapidly gaining importance. Compared to purely electronic and MEMS-based approaches,  RF processing in the photonics domain -- microwave photonics -- promises compactness and light weight, rapid tunability and integration density \cite{Capmany2007,  Marpaung2013}. Currently demonstrated optical solutions however still suffer from high RF-insertion loss and an unfavorable trade-off between achieving sufficiently narrow bandwidth, high rejection ratio and linearity.  Solutions mediated by phonons might overcome this limit as they offer a narrow linewidth without suffering from the power limits experienced in high-quality optical cavities \cite{Pant2014On-chipGeneration,Liu2018Chip-BasedSignals}

Given the high power requirements, 2D-confined waveguides lend themselves more naturally to many RF-applications. As such, stimulated Brillouin scattering (SBS) has been extensively exploited. Original work focused on phonon-photon interactions in optical fibers, which allows for high SBS-gain and high optical power but lacks compactness and integrability.  Following the demonstration of SBS-gain in integrated waveguide platforms \cite{VanLaer2015,Kittlaus2016,Pant2011}, several groups now also demonstrated RF-signal processing using integrated photonics chips.  In the most straightforward approach,  the RF-signal is modulated on a sideband of an optical carrier which is then overlayed with the narrowband SBS loss-spectrum generated by a strong pump \cite{Morrison2014}. Tuning the carrier frequency allows rapid and straightforward tuning of the notch filter over several GHz and a bandwidth below 130 MHz was demonstrated. The suppression was only 20 dB however, limited by the SBS gain achievable in the waveguide platform used, in this case a chalcogenide waveguide.  This issue is further exacerbated in more CMOS-compatible platforms, where the SBS gain is typically limited to a few dB.  This can be overcome by using interferometric approaches, which enable over 45 dB suppression with only 1 dB of SBS gain \cite{Marpaung2015,Casas-bedoya2015}.

While this approach outperforms existing photonic and non-photonic approaches on almost all specifications (see table 1 in \cite{Marpaung2015}), a remaining issue is the high RF insertion loss of about 30 dB. Integration might be  key in bringing the latter to a competitive level, as excessive fiber-to-chip losses  and high modulator drive voltages associated with the discrete photonic devices currently being used are the main origin of the low system efficiency. Also, the photonic-phononic emit-receive scheme proposed in \cite{Shin2015, Kittlaus2018RF-PhotonicOperations} results in a lower RF-insertion loss.  Although it gives up tunability, additional advantages of this approach are that its engineerable filter response \cite{Zadok2007,Shin2015} and its cascadability \cite{Kittlaus2018RF-PhotonicOperations}. Exploiting the phase response of the SBS resonance also phase control of RF signals has been demonstrated \cite{Liu2018Chip-BasedSignals}.  Both pure phase shifters and relative time delay have been demonstrated. Again interferometric approaches allow to amplify the intrinsic phase delay of the system, which is limited by the available SBS gain. In the examples above, the filter is driven by a single-frequency pump, resulting in a Lorentzian filter response.  More complex filter responses can be obtained by combining multiple pumps \cite{Thevenaz2008}.  However, this comes at the cost of the overall system response since the total power handling capacity of the system is typically limited.  As such there is still a need for waveguide platforms that can handle large optical powers and at the same time provide high SBS-gain.

Further, low-noise oscillators are also a key building block in RF-systems. Two approaches, equivalent with the two dissipation hierarchies ($\gamma \gg \kappa$ and  $\gamma \ll
\kappa$) identified in section \ref{sec:photonphononinteractions}, have been studied.  In the first case, if the photon lifetime exceeds the phonon lifetime ($\gamma \gg \kappa$), optical line narrowing and eventually self-oscillation is obtained at the transparency condition $\mathcal{C}=1$ -- resulting in substantial narrowing of the Stokes wave and thus a purified laser beam \cite{Li2012c,Suh2017Phonon-Limited-LinewidthTemperatures,Li2017MicroresonatorGyroscope}. Cascading this process leads to higher-order Stokes waves with increasingly narrowed linewidths. Photomixing a pair of cascaded Brillouin lines gives an RF carrier with phase noise determined by the lowest order Stokes wave.  Using this approach in a very low-loss silica disk resonator a phase noise suppression of 110 dBc at 100 kHz offset from a 21.7 GHz carrier was demonstrated 
\cite{Li2013a}.  In the alternate case, with the phonon lifetime exceeding the photon lifetime ($\gamma \ll
\kappa$), the Stokes wave is a frequency-shifted copy of the pump wave apart from the phase noise added by the mechanical oscillator. At the transparency condition $\mathcal{C}=1$ the phonon noise goes down, eventually reaching the mechanical Schawlow-Townes limit. Several such ``phonon lasers'' have been demonstrated already, relying on very different integration platforms \cite{Kippenberg2005,Otterstrom2018,Grudinin2010,Jiang2012High-frequencyThreshold,Morrison2017}. Further work is needed to determine if these devices can deliver the performance required to compete with existing microwave oscillators.

In the examples above, the mechanical mode is excited all-optically via a strong pump beam. Both in terms of efficiency and in terms of preventing the pump beam from propagating further through the optical circuit this may be not the most appropriate method. Recently, several authors have demonstrated electrical actuation of optomechanical circuits \cite{Fan2016, Bochmann2013c,Li2015b,Zou2016,Balram2015f,Sohn2018a,VanLaer2018}.  While this provides a more direct way to drive the acousto-optic circuit, considerable efforts are still needed to improve the overall efficiency of these systems and to develop a platform where all relevant building blocks including e.g. actuators and detectors, optomechanical oscillators and acoustic delay lines can be co-integrated without loss in performance.

\subsection{General challenges}

Each of the perspectives discussed above potentially benefits enormously from miniaturizing photonic and phononic systems in order to maximize interaction rates and pack more functionality into a constrained space. Current nanoscale electro- and optomechanical devices indeed demonstrate some of the highest interaction rates (section \ref{sec:stateoftheart}). However, the fabrication of high-quality nanoscale systems requires exquisite process control. Even atomic-scale disorder in the geometric properties can hamper device performance, especially when extended structures or many elements are required \cite{Chang2010SlowingArray,VanLaer2015c,Zhang2012c}. This can be considered the curse of moving to the nanoscale. It manifests itself as photonic and phononic propagation loss \cite{Melati2014,Patel2017a}, backscattering \cite{Melati2014}, inter-modal scattering as well as inhomogeneous broadening \cite{Wolff2016}, dephasing \cite{Sarabalis2018} and resonance splitting \cite{VanLaer2015c,Safavi-naeini2014}. To give a feel for the sensitivity of these systems, a 10 GHz mechanical breathing mode undergoes a frequency shift of about 10 MHz per added monolayer of silicon atoms \cite{VanLaer2015}. Therefore nanometer-level disorder is easily resolvable in current devices with room-temperature quality factors on the order of $10^{3}$. Developing better process control and local tuning \cite{Pfeifer2016DesignResonators} methods is thus a major task for decades to come. In addition, shrinking systems to the nanoscale leads to large surface-to-volume ratios that imply generally ill-understood surface physics determines key device properties, even with heavily studied materials such as silicon \cite{Borselli2006,Asano2017,Chan2012}. This is a particular impediment for emerging material platforms such as thin-film aluminum nitride \cite{Fan2013a}, lithium niobate \cite{Liang2017High-qualityNanocavities} and diamond \cite{Mitchell2018RealizingOptimization}. The flip side of these large sensitivities is that opto- and electromechanical systems may generate exquisite sensors of various perturbations. Amongst others, current sensor research takes aim at inertial and mass sensing \cite{Krause2012,Ekinci2004,Baker2015High-frequencyLiquids} as well as local temperature \cite{Tiebot2018ThermalNanowire} and geometry mapping \cite{Beugnot2007,Beugnot2014,Cohen2014BrillouinEmission,Zarifi2018}.

\section{Conclusion}
New hybrid electro- and optomechanical nanoscale systems have emerged in the last decade. These systems confine both photons and phonons in structures about one wavelength across to set up large interaction rates in a compact space. Similar to silicon photonics more than a decade ago, nanoscale phononic circuitry is in its infancy and severe challenges such as geometric disorder hinder its development. Still, we expect much to come in the years ahead. We believe mechanical systems are particularly interesting as low-energy electro-optic interfaces with potential use in classical and quantum information processors and sensors. Phonons are a gateway for photons to a world with five orders of magnitude slower timescales. Linking the two excitations has the potential for major impact on our information infrastructure in ways we have yet to fully explore.

\section*{Funding Information} We acknowledge the support from the U.S. government through the National Science Foundation (Grant Nos. ECCS-1509107 and ECCS-1708734) and the Air Force Office of Scientific Research under MURI. A.S.N. acknowledges the support of a David and Lucile Packard Fellowship. R.V.L. acknowledges funding from VOCATIO and from the European Union's Horizon 2020 research and innovation program under Marie Sk\l{}odowska-Curie grant agreement No. 665501 with the research foundation Flanders (FWO).

\section*{Contributions}
R.V.L. organized and wrote much of the manuscript along with A.S.N. The section on microwave signal processing was written by D.V.T. All authors read, discussed and gave critical feedback on the paper.

\bibliography{Mendeley_Optica_review}

\appendix

\section{Theoretical description}
\label{sec:appendix_theory}
We first give a description of 3D-confined cavity optomechanics. Next, we connect to the description of 2D-confined waveguides.

\subsection{Cavities: 3D-confined}
\label{subsec:appendix_theory_cavities}

We focus on cavity optomechanics in this section although much of it applies to electromechanics as well. The dynamics of an optomechanical system involves taking into account the interplay between the coupled acoustic and optical degrees of freedom in a system. The frequencies of these two coupled degrees of freedom are typically different by many orders of magnitude so that the only physical significant coupling arises parametrically in the form described below.  As a first step, a modal decomposition of time varying deformations in the elastic structure is considered, so a set of parameters $x_j(t)$, each encoding the deformation due to a particular vibrational mode is considered. A similar decomposition of Maxwell's equations leads to a set of electromagnetic modes of the structure with amplitudes $\alpha_j(t)$, which with the correct normalization would lead to $U_j = \hbar \omegao |\alpha_j|^2$ being the energy and $|\alpha_j|^2$ the average photon number in mode $j$. Each optical (mechanical) mode of the structure has a frequency $\omegao$ ($\omegam$) and their associated dynamics. At first we will only consider the interaction between two modes: a single optical and a single mechanical mode of the structure. Optomechanical interactions give rise to coupling between these modes in the following way: the deformation of the structure in a specific vibrational mode parametrized by $x$, causes a change in the optical frequency given by $\omegao(x) = \omegao(0) + G x$, where the optomechanical coupling parameter $G = \partial_{x}\omegao$ has units of $\text{Hz}\cdot\text{m}^{-1}$. The modal equation for the electromagnetic field, under laser excitation at frequency $\omegain = \omegao(0) - \Delta$ with input photon flux given by $|\alpha_\text{in}|^2$ is then expressed as
\begin{eqnarray}
\frac{d \alpha(t)}{dt} = -\left(i(\Delta + G x) + \frac{\kappa}{2} \right) \alpha - \sqrt{\kappaex} \alpha_\text{in}.\label{eqn:optical1}
\end{eqnarray}
The cavity decay rate $\kappa$ represents the full-width half-maximum (FWHM) of the optical mode excitation spectrum and contains all decay channels coupling to the photonic system. Typically, $\kappa$ consists of an engineered \textit{extrinsic} decay rate $\kappaex$ as well as the \textit{intrinsic} loss rate $\kappain$. 

\subsubsection{Linear detection of motion}\label{subsubsec:linear_detection}
First, we consider how motion is detected optically in such a setup, completely ignoring at first the effect of the light on the mechanical system. We make a few approximations for this particular analysis that are useful though not generally valid. First we assume that $x(t)$ is slow compared to optical bandwidth $\kappa$, or equivalently $\omegam \ll \kappa$. Also we assume that the laser field is driving the photonic cavity on resonance so $\Delta = 0$, and that the optical decay rate is dominated by the out-coupling so $\kappa = \kappaex$. The output field is then given by $\alpha_\text{out} = \alpha_\text{in} + \sqrt{\kappa} \alpha$, which for oscillations that are small, i.e. when the laser-cavity detuning is being modulated by the motion within the linear region of the cavity phase response such that $G x \ll \kappa$, can be solved to obtain a relation representing phase modulation of the output field: $\alpha_\text{out} = -\alpha_\text{in}(1 - i 4 G x(t) /\kappa)$. This is a first order expansion of $e^{-i\phi(t)}$ where $\phi(t) = G x(t)/\kappa$. An alternate way of writing this expression is in terms of the intracavity field $\alpha(t)$ which, neglecting the mechanical motion, is given by $\alpha = - 2\alpha_\text{in}/\sqrt{\kappa}$. The output field is then
\begin{eqnarray}
\alpha_\text{out} = -\alpha_\text{in}  - i \sqrt{\Gamma_\text{meas}} x/\xzp \label{eqn:aout_ain_x}
\end{eqnarray}
with the measurement rate  defined as 
\begin{eqnarray}
\Gamma_\text{meas} \equiv \frac{4 G^2 \xzp^2 |\alpha|^2}{\kappa} = \frac{4 g^{2}_{0}|\alpha|^{2}}{\kappa} \label{eqn:Gamma_meas}
\end{eqnarray}
being the rate at which photons are scattered from the laser beam to sidebands due to motion of amplitude $\xzp$ and an intracavity optical field intensity of $|\alpha|^2$. The subscript represents \textit{zero-point}, assigned in anticipation of the quantum analysis below though for a classical description $\xzp$ can be used to normalize the above expression without changing the physics. The measurement rate as defined here is central to understanding the operation of optomechanical systems in the linear regime and will be used throughout the text below often denoted alternatively as $\gamma_\text{OM} \equiv \Gamma_\text{OM} \equiv \Gamma_\text{meas}$. The zero-point coupling rate $g_{0}=G \xzp$ defined here is consistent with section \ref{sec:photonphononinteractions} in the main text.

\subsubsection{Back-action on the mechanical mode} Now we consider how the motion of the mechanical system is modified due to interaction with the optical field resonating in the structure. In addition to equation \ref{eqn:optical1}, to understand the back-action arising from the interplay between the optical field and mechanical motion, we must consider the dynamics of the motional degree of freedom:
\begin{eqnarray}
\ddot{x}(t) +  \gammain \dot{x} + \omegam^2 x  =( F_\text{BA}(t) + F_\text{input}(t) )/ \meff. \label{eqn:mech1}
\end{eqnarray}
The left-hand side of the above equation is simply the equation of motion for a damped harmonic oscillator and takes into account the dynamics of the modal degree of freedom being considered. The right-hand side of the equation are the forcing terms: $F_\text{BA}(t)$ is the optical back-action, while $F_\text{input}(t)$ is an input force which we use to understand the linear response of the mechanical system. The back-action force is given by radiation pressure described via the Maxwell stress tensor, which is quadratic in the field or proportional to $|\alpha|^2$. By considering the total energy of the system (see section~\ref{subsec:theory_quantum}), we find that $F_\text{BA}(t) = \hbar G |\alpha(t)|^2$.  Equations \ref{eqn:optical1} and \ref{eqn:mech1} now describe the dynamics of the coupled system and can be solved to obtain the effects of back-action in the classical domain. In particular, we are primarily interested in the modification of the linear response of the mechanical system to an input force,  i.e. changes to its damping rate and frequency. These changes come about from the mechanical motion $x$ modifying the intracavity field $\alpha(t)$ which then applies a force back onto $x$ which can be proportional, lagging, or leading, leading to a redefinition of the mechanical system's complex frequency. To calculate the laser power and frequency dependence of these modifications, we choose an operating point $(\mean{\alpha},\mean{x})$ and linearize the equations of motion by taking to account only the dynamics of the fluctuations $\delta x(t)$ and $\delta \alpha(t)$. This gives us a set of three coupled linear differential equations
\begin{eqnarray}
\delta \ddot{x}(t) &=& -\omegam^2 \delta x - \gammain \delta \dot{x} +\nonumber\\ &&\hbar G(\mean{\alpha}^\ast \delta \alpha + \text{c.c.})/\meff + F_\text{input}(t)/\meff \label{eqn:om1x}\\
\delta \dot{\alpha}(t) &=& -(i\Delta + \kappa/2)\delta \alpha - iG \mean{\alpha} \delta x - \sqrt{\kappaex} \delta \alpha_{\text{in}} \label{eqn:om2x} \\
\delta \dot{\alpha}^\ast(t) &=& (i\Delta - \kappa/2)\delta \alpha^\ast + iG \mean{\alpha}^\ast \delta x - \sqrt{\kappaex}\delta \alpha^{\star}_{\text{in}} \label{eqn:om3x}
\end{eqnarray}
which can be solved for input forces $F_\text{input}(t)$ taking $\delta \alpha_{\text{in}}=0$ for now. Solving these equations in the Fourier domain, we obtain an expression for the small-signal response of the mechanical system to the input force in terms of a dispersion relation, $\delta x(\omega) \equiv \chi_x(\omega) F_\text{input}(\omega)$, with
\begin{eqnarray}
\chi_x(\omega) = \frac{1}{\meff(\omegam^2 - \omega^2 - i\omega\gammain + \Sigma_\text{opt}(\omega))},\label{eqn:mech_response}
\end{eqnarray}
where
\begin{eqnarray}
\Sigma_\text{opt}(\omega) &=& -i\hbar G^2 |\mean{\alpha}|^2\left(\chi_\alpha(\omega) - \chi_\alpha^\ast(-\omega)\right)/\meff
\end{eqnarray}
and $\chi_\alpha(\omega) = (i(\Delta - \omega) + \kappa/2)^{-1}$ is the optical resonance response function. The expression in equation \ref{eqn:mech_response} represents the response of a damped mechanical resonance that is modified by a ``self-energy'' term, $\Sigma_\text{opt}(\omega)$, due to interaction with optical resonance. The real and imaginary parts of this self-energy cause an effective modification of the mechanical frequency and linewidth $\omegam$ and $\gammain$. This shift in the complex frequency, often referred to as the ``optical spring'' and ``optical damping/amplification'' effects can be expressed succinctly in terms of $\Sigma_\text{opt}(\omega)$:
\begin{eqnarray}
\gamma_\text{OM} &\approx& -\frac{\text{Im} \Sigma_\text{opt}(\omegam)}{\omegam},~\text{and}\\
~~~\delta\omegam &\approx& \frac{\text{Re} \Sigma_\text{opt}(\omegam)}{2\omegam}.
\end{eqnarray}
These expression are good approximations in the weak-coupling regime ($g\ll \kappa + \gamma$). In the strong-coupling regime ($g\gg \kappa + \gamma$), the full frequency-dependence of the self-energy $\Sigma_\text{opt}(\omega)$ should be considered.

A common mode of operation of optomechanical systems that are sideband-resolved ($\omegam \gg \kappa$) is to tune the laser approximately a mechanical frequency to the red side of the optical resonance so $\Delta = \omegam$. In this case, the above relations lead us to $\gamma_\text{OM} \approx 2\hbar G^2 |\mean{\alpha}|^2/(\kappa\meff\omegam)=4  G^2 \xzp^2 |\mean{\alpha}|^2/\kappa$ which is seen to be equal to the measurement rate calculated in equation \ref{eqn:Gamma_meas}, though that was for a different regime of operation. The equality of these two rates can be understood as such: with the red-detuned scheme of driving, all of the sideband scattering, which occurs at rate $\Gamma_\text{meas}$, causes up-shifting of the laser photons into the photonic mode, and thus effectively damps the mechanical resonator's motion. In the above we focused on the effect of the optomechanical interaction on the mechanical resonator's response function. However, there are equally important changes in the electromagnetic response. These effects, including Brillouin gain/loss and slow/fast light, can be derived similarly \cite{VanLaer2016}.

\subsubsection{Understanding optomechanical coupling in a nanophotonic system}
\label{subsec:theory_quantum}

In the previous section we studied how optomechanical coupling can be used to detect mechanical motion and modify the linear response of a mechanical resonator. Here we will see how such an interaction comes about in a realistic nanophotonic system. Though a toy model with a one-dimensional scalar wave-equation and a simplified mass-spring system has long been used in studying optomechanical systems, obtaining a precise understanding of the coupling rates given nontrivial wavelength-scale optical and elastic mode profiles requires careful consideration of the fields and calculation of the interactions. The goal of this section is to show how we can obtain equations similar to equations \ref{eqn:om1x}-\ref{eqn:om3x} where $x(t)$ and $\alpha(t)$ now represent mode amplitudes for acoustic and optical excitations in a nanophotonic device.

We start by solving separately the dynamical equations for electromagnetics and elastodynamics which can be expressed as eigenvalue equations for the magnetic field $\m H \fr$ and elastic displacement field $\m Q \fr$ respectively:
\be
\mathsf{L} \m h_j = \omega_j^2 \m h_j,\qquad \mathsf{L}(\cdot) = c^2 \curl\left[ \frac{\epsilon_0}{\dyad \epsilon \fr}\curl(\cdot)
\right]. \label{eqn:eigen}
\ee
\begin{eqnarray}
\omega^2_j \m Q_j (\m r) &=& \mathsf{L} \m Q_j(\m r)\nonumber\\
&&\mathsf{L}(\cdot) = -\frac{\lambda+\mu}{\rho} \nabla(\nabla \cdot ( \cdot )) - \frac{\mu}{\rho} \nabla^2(\cdot).
\end{eqnarray}
The set of solutions of these two equations are the normal electromagnetic and acoustic modes of the structure, and define the spectrum. Typically, a software package such as COMSOL is used to obtain these solutions in dielectric structures that don't permit analytic analysis. Valid solutions of the electromagnetic and  elastic field in the structure can then be expressed as normal mode expansions
\begin{eqnarray}
\op{\m E}{} (\m r, t) &=& \sum_j \m e_j \fr a_{j}(t)+ \text{h.c.},
\end{eqnarray}
\begin{eqnarray}
\op{\m Q}{}  (\m r, t) &=& \sum_k \m Q_k \fr \op{b}{k}(t)+ \text{h.c.}
\end{eqnarray}
with $\dot{a{}}_j(t) = -i \omega_j a_{j}(t)$, and $\dot{b}_k(t) = -i \omega_k \op{b}{k}(t)$. In defining the quantum field theory, we assign to each mode $j$ a Hilbert space $\{\ket{n}_j,n=0,1,2,3,..\}$ where $\ket{n}_j$ is the state representing $n$ photons or phonons in the $j$th mode. Phonons and photons in each of these Hilbert spaces are then annihilated with the operators $a_{j}$ and $b_{j}$ respectively. The normalizations of $\m e_j$ and $\m Q_k$ in the equations above are then physically significant since, \eg, the expectation value of $\hbar\omega_j a^{\dagger}_{j}a_{j}$ represents the energy stored in the $j$th mode of the electromagnetic field. We can use the classical expressions for energy in the fields to calculate the energy for a single photon/phonon above the vacuum state, $\ket{\psi} = \ket{1}_j$, which we will then set equal to $\hbar\omega_j$:
\begin{eqnarray}
\hbar\omega_j \equiv U^{\ket{\psi}}_\text{mech} &=& \bra{\psi}\int \text{d}\m r~ \dot{\op{\m Q}{}} \fr \rho \fr \dot{\op{\m Q}{}} \fr \ket{\psi} \nonumber\\&&- \bra{\text{vac}}\int \text{d}\m r~  \dot{\op{\m Q}{}} \fr \rho \fr \dot{\op{\m Q}{}} \fr \ket{\text{vac}} \nonumber\\
 &=& 2 \omega_j^2 \int \text{d}\m r~ \m Q^\ast_j \fr \rho \fr \m Q_j \fr \nonumber \\
 &=& 2 m_\text{eff} \omega_j^2 \text{max} [ |\m Q_j \fr |^2].
\end{eqnarray}
where we've defined the effective mass and zero-point motion of the mode to be
\begin{eqnarray}
m_{\text{eff},j} = \frac{\int \text{d}\m r~  \m Q^\ast_j \fr \rho \fr \m Q_j \fr}{\text{max}[  |\m Q_j \fr |^2]},~~~\text{and}~~~~\nonumber\\x_{\text{zpf},j} \equiv \text{max} [ |\m Q_j \fr| ]  = \sqrt{\frac{\hbar }{2 m_{\text{eff},j} \omega_j}}.\label{eqn:x_zpf}
\end{eqnarray}
A similar consideration for the electromagnetic fields leads to
\begin{eqnarray}
\hbar\omega_j \equiv U_\text{em} &=& \bra{\psi}\int \text{d}\m r~ \op{\m E}{} \fr \dyad \epsilon \fr \op{\m E}{} \fr \ket{\psi} - \nonumber\\ && \bra{\text{vac}}\int \text{d}\m r~ \op{\m E}{} \fr \dyad \epsilon \fr \op{\m E}{} \fr \ket{\text{vac}} \nonumber\\
 &=& 2 \int \text{d}\m r~ \m e^\ast_j \fr \dyad \epsilon \fr \m e_j \fr \nonumber\\
 &=& 2 V_\text{eff} \text{max} [ \m e^\ast_j \fr \dyad \epsilon \fr \m e_j \fr ].
\end{eqnarray}
with
\begin{eqnarray}
V_{\text{eff},j} = \frac{\int \text{d}\m r~ \m e^\ast_j \fr \dyad \epsilon \fr \m e_j \fr }{\text{max}[ \m e^\ast_j \fr \dyad \epsilon \fr \m e_j \fr ]}.
, ~~~\text{and}~~~~\nonumber\\\text{max} [ |\m e_j \fr| ]  = \sqrt{\frac{\hbar \omega_j}{2 V_{\text{eff},j} \epsilon_\text{diel}}}.
\end{eqnarray}

Having defined the quantization of the fields and mode normalizations, we can now write a Hamiltonian for the optomechanical system,
\begin{eqnarray}
\mathcal{H} = \underbrace{\hbar \sum_j \omega_j a^{\dagger}_{j}a_{j}}_{\mathcal{H}_{\text{o}}}+\underbrace{\hbar \sum_j \omega_j b^{\dagger}_{j}b_{j}}_{\mathcal{H}_{\text{m}}} + \mathcal{H}_{\text{int}}
\end{eqnarray}
that can capture the quantum dynamics. The challenge remains calculation of the optomechanical interaction term $\mathcal{H}_{\text{int}}$. We are interested in interactions of the type $a^{\dagger}_{j}a_{k}({b}_{l}+{b}^\dagger_{l})$ which are the simplest type that allow energy conservation, assuming that the photonic frequencies for mode $j$ and $k$ are many orders of magnitude larger than the mechanical frequency of mode $l$. For the case where $j=k$, we are considering a shift in the frequency of optical mode $j$ due to a mechanical displacement in mode $l$. The relevant interaction energy or rate can be calculated using first-order cavity perturbation theory. In a dielectric structure characterized by $\dyad \epsilon_0 \fr$, modifications due to deformations can be taken into account with the expression
\be
\dyad \epsilon \fr = \dyad \epsilon_0 \fr + \dyad{\delta \epsilon}\fr.
\ee
To first order, such a modification of the dielectric causes a shift in the optical resonance frequency of a mode with mode profile $\m e\fr$ of
\begin{eqnarray}
\omega^{(1)} = -\frac{\omega_0}{2}\frac{\langle \m e | \dyad{\delta \epsilon} | \m e \rangle}{\langle \m e |\dyad{ \epsilon}| \m e \rangle}.\label{eqn:basic_perturbation_result}
\end{eqnarray}
There are two ways that the dielectric constant changes due to this deformation. First, a deformation of the optical resonator affects the dielectric tensor at the \emph{boundaries} between different materials. This is because the high-contrast step profile of $\dyad{ \epsilon}\fr$ across a boundary is shifted by deformations of the structure. By relating a deformation to a change in the dielectric constant, we can use equation~\ref{eqn:basic_perturbation_result} to calculate the optomechanical coupling. Johnson has derived a useful expression~\cite{Johnson2002} for this shift in frequency, which when adapted to optomechanics~\cite{Eichenfield2009}, gives a frequency shift per unit displacement of
\begin{eqnarray}
g_\text{OM,B} = -\frac{\omega_0}{2} \frac{\int Q_n(\m r)(\Delta \dyad{\epsilon} |\m e^\parallel |^2 - \Delta(\dyad{\epsilon^{-1}})|\m d^\perp|^2) \text{d}A}{ \text{max}(|\m Q|) \int \dyad{\epsilon}(\m r)|\m e(\m r)|^2 \text{d}^3\m r} \label{eqn:gOMBND}
\end{eqnarray}
for a mechanical vector displacement field $\m Q \fr$ with component $Q_n \fr$ normal to the boundary. Secondly, a \emph{photoelastic} contribution to the optomechanical coupling arises from local changes in the refractive index due to stress in the material induced by the mechanical deformation. For a particular displacement vector $\m Q \fr$  corresponding strain tensor $S_{ij} = \frac{1}{2} \left( \partial_i  Q_j + \partial_j  Q_i \right)$, the dielectric perturbation is given by
\begin{eqnarray}
\dyad{\delta \epsilon}\fr = \dyad{\epsilon} \cdot \frac{\dyad{p} \cdot \dyad{S}}{\epsilon_0} \cdot \dyad{\epsilon},
\end{eqnarray}
which reduces to $\delta \epsilon_{ij} = - \epsilon_0 n^4 p_{ijkl} S_{kl}$ for an isotropic medium. The fourth-rank tensor $\dyad{p}$ with components $p_{ijkl}$ is called the photoelastic tensor and is a property of the material. The \textit{roto-optic} effect, which captures permittivity changes induced by rotation, must be included as well in optically anisotropic materials \cite{Nelson1970NewScattering,Weis1985LithiumStructure,Smith2017EnhancedMedia}. Composite metamaterials may yield enhanced photoelasticity \cite{Smith2017EnhancedMedia}. Often, when considering the symmetries in the atomic structure of the material, a reduced tensor is used with elements $p_{ij}$. The perturbation integral can then be used to calculate the shift in frequency per unit displacement:
\begin{eqnarray}
g_\text{OM,PE} = -\frac{\omega_0}{2} \frac{\int ~\m e \cdot \dyad{\delta \epsilon} \cdot \m e ~\text{d}^3\m r }{ \text{max}(|\m Q|) \int \dyad{\epsilon}(\m r)|\m e(\m r)|^2 \text{d}^3\m r}.\label{eqn:gOMPE}
\end{eqnarray}
These expressions give the boundary and photoelastic components for a shift in the optical cavity frequency per unit displacement of the maximum deflection point of a deformation profile $\m Q \fr$. A natural unit for displacement is the zero-point fluctuation amplitude found by multiplying the expressions~(\ref{eqn:gOMBND}) and (\ref{eqn:gOMPE}) by the zero-point fluctuation length $x_{\text{zpf}} = \sqrt{\hbar/(2 m_{\text{eff}} \omegam)}$ (see equation~\ref{eqn:x_zpf}). The  coupling rate is $g_0 = g_{0,\text{Bnd}} + g_{0,\text{PE}}$, and the corresponding optomechanical interaction Hamiltonian can be written as
\begin{eqnarray}
\mathcal{H}_{\text{int}} &=& \hbar (g_\text{OM,PE} + g_\text{OM,Bnd})x a^{\dagger} a{}\nonumber\\
&=& \hbar g_0 (b^{\dagger} + b) a^{\dagger} a.
\end{eqnarray}

\subsection{Waveguides: 2D-confined}
A theory for 2D-confined waveguides with translational symmetry can be developed similarly to the previous section on 3D-confined cavities. We refer to \cite{Sipe2016,Wolff2015,Zoubi2016b} for a thorough treatment from first principles. Here, we focus on connecting the waveguides' dynamics to the cavities' dynamics described in the previous section. For a waveguide the translational symmetry implies that the photonic and phononic eigenproblems can be expressed in terms of wavevectors $\beta$ and $K$, yielding as solution dispersion relations $\omega(\beta)$ and $\Omega(K)$. As shown in \cite{Sipe2016,Wolff2015,Zoubi2016b}, the Hamiltonian of the waveguide is $\mathcal{H} = \mathcal{H}_{\text{free}}+\mathcal{H}_{\text{int}}$ with the free Hamiltonian given by
\begin{equation}
\label{eq:freeHamil}
\mathcal{H}_{\text{free}} = \int \text{d}\beta \, \hbar \omega_{\beta} a^{\dagger}_{\beta}a_{\beta} + \int \text{d}K \, \hbar \omega_{K} b^{\dagger}_{K}b_{K}
\end{equation}
and the interaction Hamiltonian given by
\begin{equation}
\label{eq:linearHamilwg2}
\mathcal{H}_{\text{int}} = \frac{\hbar}{\sqrt{2\pi}} \int \int \text{d}\beta\,\text{d}K \, \left(g_{\beta + K} a_{\beta} b_{K}  + g_{\beta - K} a_{\beta} b^{\dagger}_{K} + \text{h.c.}\right)
\end{equation}
in the momentum-description where the three-wave mixing interaction rate $g_{\beta \pm K} = g_{0 | \beta \pm K} \alpha^{\dagger}_{\beta \pm K}$ is proportional to the pump amplitude $\alpha_{\beta \pm K}$ of the mode with wavevector $\beta \pm K$. Here we assume phase-matching ($\Delta K = -\betap + \beta \pm K = 0$ with $\betap$ the pump wavevector) and usually consider $\alpha$ to represent a strong pump that can be treated classically.

Previous work \cite{VanLaer2016} linked the waveguide's coupling rate $g_{0|\beta \pm K}$ and Brillouin gain coefficient $\mathcal{G}_{\text{B}}$ to an optomechanical cavity's coupling rate $g_{0}$ via a mean-field transition performed on the photonic and phononic equations of motion both in the limit of large and small cavity finesse. Here, we derive the same connection but now via a mean-field transition in the large-finesse limit performed directly on the waveguide's Hamiltonian given by expression \ref{eq:linearHamilwg2}. We consider a cavity of roundtrip length $L$ constructed from a waveguide described by equation \ref{eq:linearHamilwg2} and focus on a triplet of two photonic and one phononic mode(s). The operator corresponding to the number of excitations in each of the modes can be expressed as
\begin{align}
 c^{\dagger} c  &= \int_{k_{c} - \pi/L}^{k_{c}+\pi/L} \, \text{d}k \, c^{\dagger}_{k}c_{k} \\ \notag
& = \frac{2\pi}{L} c^{\dagger}_{k_{c}}c_{k_{c}} 
\end{align}
with $k_{\text{c}}$ a relevant center wavevector. Therefore, the cavity and waveguide operators are linked by
\begin{equation}
c = \sqrt{\frac{2\pi}{L}} \, c_{k_{c}}
\end{equation}
with $c$ either a photonic or phononic ladder operator. Considering the first term in $\mathcal{H}_{\text{int}}$, we therefore have
\begin{align}
 \int \int \text{d}\beta\,\text{d}K \, g_{\beta + K} a_{\beta} b_{K} &= \left( \frac{2\pi}{L}\right)^{2} g_{\beta + K} a_{\beta} b_{K} \\
 &= \frac{2\pi}{L} g_{0|\beta+K} \alpha_{\beta+K} \delta a \delta b \\
 &= \frac{g_{0|\beta+K}}{\sqrt{L}} \alpha \delta a \delta b \\ 
 & = g_{0} \alpha \delta a \delta b
\end{align}
with $\delta a$ and $\delta b$ the cavity's photonic and phononic ladder operators. Thus we obtain
\begin{equation}
g_{0} = \frac{g_{0|\beta+K}}{\sqrt{L}}
\end{equation}
as in equation \ref{eq:linkg0} in the main text. This link between the coupling rates holds both for forward and backward as well as for intra- and inter-modal scattering \cite{VanLaer2016}.

Next, we consider the waveguide's dynamics in slightly more detail. The dynamics of an optomechanical waveguide is usually considered after transforming from momentum- to real-space operators
\begin{equation}
c(z) = \int \, \frac{\text{d}k}{2\pi} \, e^{-i(k - k_{\text{c}})z} c_{k}
\end{equation}
We are usually concerned with narrow bandwidths relative to the group-velocity-dispersion so the frequencies can be expanded to first-order as
\begin{equation}
\label{eq:expandbands}
\omega_{k} \approx \omega_{\text{c}} + \vg (k - k_{\text{c}})
\end{equation}
A few manipulations \cite{Sipe2016} starting from equation \ref{eq:freeHamil} lead to
\begin{equation}
\label{eq:freespatial}
\mathcal{H}_{\text{free}} =  \hbar \int \, \text{d}z \, \left[\, a^{\dagger}(z) \hat{\omega}_{a} a(z)  + \, b^{\dagger}(z) \hat{\omega}_{\text{m}} b(z) \right]
\end{equation}
with $\hat{\omega}_{k} =  \omega_{\text{c}} - i \vg \partial_{z}$ the real-space operator corresponding to the momentum-space dispersion relation $\omega_{\text{k}}$. Higher-order expansions of the dispersion relation in equation \ref{eq:expandbands} yield higher-order spatial derivatives the operator $\hat{\omega}_{k}$.

Further, dropping the second term in equation \ref{eq:linearHamilwg2} the real-space interaction Hamiltonian becomes
\begin{equation}
\label{eq:interactionspatial}
\mathcal{H}_{\text{int}} = \hbar\int \, \text{d}z \, g_{0|\beta+K} \alpha^{\dagger}(z) a(z) b (z) + \text{h.c.}
\end{equation}
Further, the Heisenberg equations of motion $\dot{c}(z) = -\frac{i}{\hbar}[c(z),\mathcal{H}_{\text{int}}]$ together with the equal-time commutator $[c(z),c^{\dagger}(z')] = \delta(z-z')$ yield
\begin{align}
\label{eq:motionwg}
\notag
(\partial_{t} + v_{\alpha} \partial_{z}) \alpha &= -i\omega_{\alpha} \alpha - i g_{0|\beta + K} a \, b \\
(\partial_{t} + v_{a} \partial_{z}) a &= -i\omega_{a} a - i g_{0|\beta+K}\alpha \, b^{\dagger} \\ \notag
(\partial_{t} + \vm \partial_{z}) b &= -i\omegam b - i g_{0|\beta+K}\alpha \, a^{\dagger}
\end{align}
These equations describe the spatiotemporal evolution of the three interacting fields in absence of dissipation and phase-mismatch. Intrinsic propagation losses can be included via a dissipative term, e.g.
\begin{equation}
(\partial_{t} + \vm \partial_{z}) b = -i\omegam b - \kappam b - i g_{0|\beta+K}\alpha a^{\dagger}
\end{equation}
for the phononic field with $\kappam = \vm \alpham $ with $L_{\text{m}} = \alpham^{-1}$ the phononic decay length. Similarly, a non-zero phase-mismatch $\Delta K \neq 0$ effectively reduces the interaction rate. In particular, dropping the second term for simplicity the momentum-space interaction Hamiltonian \ref{eq:linearHamilwg2} becomes
\begin{equation}
\label{eq:linearHamilwg3}
\mathcal{H}_{\text{int}} = \frac{\hbar}{\sqrt{2\pi}} \int \int \int \text{d}\beta\,\text{d}K \, \text{d}\betap \left(g_{0|\betap}\alpha^{\dagger}_{\betap} a_{\beta} b_{K}  L \, \text{sinc}\left(\frac{\Delta K L}{2}\right)  + \text{h.c.}\right)
\end{equation}
with $L$ the waveguide length. Thus the finite length weakens the wavevector-selectivity, generating interactions between a larger set of modes. The strongest interactions are obtained between modes for which $\Delta K = 0$. This corresponds to the real-space Hamiltonian
\begin{equation}
\label{eq:interactionspatial2}
\mathcal{H}_{\text{int}} = \hbar\int \, \text{d}z \, g_{0|\beta+K} \alpha^{\dagger}(z) a(z) b (z) \, e^{i\Delta K_{\text{c}} z} + \text{h.c.}
\end{equation}
with $\Delta K_{\text{c}} = -\beta_{\text{pc}} + \beta_{\text{c}} + K_{\text{c}}$ the phase-mismatch between the center wavevectors of the photonic and phononic fields. This suppresses the interaction in the equations of motion via a rotating term, for instance
\begin{equation}
(\partial_{t} + \vm \partial_{z}) b = -i\omegam b - i g_{0|\betap}\alpha a^{\dagger} \, e^{-i \Delta K_{\text{c}} z}
\end{equation}
The range of spatiotemporal effects described by these and extended versions of these equations of motion are considered in detail in amongst others \cite{VanLaer2016,Kang2009,Wolff2015,Sipe2016,Wolff2015bb,Rakich2018,Renninger2018,Koehler2017,Butsch2012}.

\section{Single-photon nonlinearity}
In this section we give a derivation for relationships \ref{eq:wgcrossKerr} and \ref{eq:finnesecavwg}. We consider a 2D-confined waveguide of length $L$ and inject a photon flux $\langle\Phi\rangle=\vg/L$ with $\vg$ the optical group velocity that corresponds to one photon on average in the waveguide. This photon excites the mechanical system with a displacement $x_{1}$, which in turn yields a phase-shift $\vartheta_{\text{wg}}$ on a second probe photon. The phase-shift can be expressed as
\begin{equation}
\vartheta_{\text{wg}} = k_0 (\partialx \neff) x_{1} L 
\end{equation}
with $k_0$ the vacuum optical wavevector and $\partialx \neff$ the sensitivity of the effective optical index to mechanical motion $x_{1}$. Assuming a static displacement, we have $x_{1} = \langle F\rangle/\keff$ with $\langle F \rangle$ the force exerted by the first photon and $\keff$ the effective mechanical stiffness per unit length. In addition, from power-conservation it can be shown that $\langle F\rangle = \frac{1}{c} \partialx \neff (\hbar \omega)\langle\Phi\rangle$ \cite{Rakich2009}. Substitution leads to
\begin{equation}
\vartheta_{\text{wg}} = \frac{\hbar \omega^{2}}{\keff} \left(\frac{1}{c}\partialx \neff\right)^{2} \vg
\end{equation}
Since the Brillouin gain coefficient can be expressed as \cite{VanLaer2015}
\begin{equation}
\mathcal{G}_{\text{B}} = 2 \omega \frac{\Qm}{\keff} \left(\frac{1}{c}\partialx \neff\right)^{2}
\end{equation}
this yields
\begin{equation}
\vartheta_{\text{wg}} = \frac{\hbar \omega}{2} \frac{\mathcal{G}_{\text{B}}}{\Qm} \vg
\end{equation}
Making use of equation \ref{eq:linkcavwg} gives
\begin{equation}
\vartheta_{\text{wg}} = \frac{2 g^{2}_{0|\beta + K}}{\vg \omegam}
\end{equation}
This is the single-photon cross-phase shift for a statically driven mechanical waveguide. When the mechanics is driven with a detuning $\Delta\Omega > \gamma$ close to the resonance, one must replace $x_{1} \rightarrow (\omegam/(2\Delta\Omega)) x_{1}$ so the cross-Kerr phase shift increases to
\begin{equation}
\vartheta_{\text{wg}} = \frac{g^{2}_{0|\beta + K}}{\vg \Delta \Omega}
\end{equation}
which is in agreement with more rigorous analysis \cite{Zoubi2017}. This phase-shift is independent of the waveguide length $L$ since the photon flux and the optical forces are inversely proportional to length $L$ when there is on average a single photon in the waveguide.

Next, we link the waveguide single-photon phase-shift $\vartheta_{\text{wg}}$ to the cavity single-photon phase shift
\begin{equation}
\vartheta_{\text{cav}} = \frac{2g_0^{2}}{\kappa \Delta \Omega}
\end{equation}
derived in section \ref{sec:perspectives}\ref{subsec:singlephotonNLO} of the main paper. Making use of equation \ref{eq:linkg0}, the cavity phase-shift $\vartheta_{\text{cav}}$ can be expressed as
\begin{equation}
\vartheta_{\text{cav}} = \frac{2g^{2}_{0|\beta+K}}{ L_{\text{rt}}\kappa \Delta \Omega}
\end{equation}
with $L_{\text{rt}}$ the cavity roundtrip length. The cavity finesse is $\mathcal{F} = 2\pi/(\kappa T_{\text{rt}})$ with $T_{\text{rt}} = L_{\text{rt}}/\vg$ the cavity roundtrip time so we obtain
\begin{equation}
\vartheta_{\text{cav}} = \frac{\mathcal{F}}{\pi} \vartheta_{\text{wg}}
\end{equation}
The expression for the cavity phase-shift assumed critical coupling and a small phase shift given by $\vartheta_{\text{cav}} = 2\Delta/\kappa$ with $\Delta = \vartheta_{\text{wg}}/T_{\text{rt}}$ the mechanically-induced detuning from the cavity resonance.

\end{document}